\documentclass[journal]{IEEEtran}
\usepackage{amsmath,amssymb,amsfonts}
\usepackage[ruled,vlined,linesnumbered]{algorithm2e}
\usepackage{algpseudocode}
\usepackage{xr}
\usepackage{pifont}
\usepackage{tabularx}
\usepackage{cellspace}
\makeatletter
\newcommand{\removelatexerror}
{\let\@latex@error\@gobble}
\makeatother
\usepackage{graphicx}
\usepackage{textcomp}
\usepackage{xcolor}
\usepackage{array}
\usepackage{textcomp}
\usepackage{dutchcal}
\usepackage{multirow}
\usepackage{bm}
\usepackage{booktabs}
\usepackage{ntheorem}
\usepackage{subfigure} 
\usepackage{orcidlink} 
\hypersetup{hidelinks}
\usepackage[mathscr]{euscript}
\usepackage{euscript}
\usepackage{mathtools}
\usepackage{epstopdf}
\usepackage{cases}
\def\BibTeX{{\rm B\kern-.05em{\sc i\kern-.025em b}\kern-.08em
		T\kern-.1667em\lower.7ex\hbox{E}\kern-.125emX}}
\usepackage{cases}
\theoremseparator{.}

\theorembodyfont{}

\definecolor{color}{rgb}{0, 0, 0}
\definecolor{color_revise}{rgb}{0, 0, 0}

\begin{document}

\title{Active RIS-Empowered Covert Satellite-Terrestrial Communications}

\author{Chuang~Zhang$^{\orcidlink{0000-0002-0505-0512}}$,\IEEEmembership{}
        Geng~Sun$^{\orcidlink{0000-0001-7802-4908}}$,~\IEEEmembership{Senior Member,~IEEE,}
        Jiahui~Li$^{\orcidlink{0000-0002-7454-3257}}$,\IEEEmembership{}
        Shiwen~Mao$^{\orcidlink{0000-0002-7052-0007}}$,~\IEEEmembership{Fellow,~IEEE,}
        and Abbas~Jamalipour$^{\orcidlink{0000-0002-1807-7220}}$,~\IEEEmembership{Fellow,~IEEE}
        
        \thanks{
        \par Chuang Zhang, Geng Sun and Jiahui Li are with the College of Computer Science and Technology, Key Laboratory of Symbolic Computation and Knowledge Engineering of Ministry of Education, Jilin University, Changchun 130012, China. (e-mails: chuangzhang1999@gmail.com, sungeng@jlu.edu.cn, lijiahui@jlu.edu.cn).
        \par Shiwen Mao is with the Department of Electrical and Computer Engineering, Auburn University, Auburn, AL 36849, USA (e-mail: smao@ieee.org).
        \par {\color{black}{Abbas Jamalipour is with the School of Electrical and Computer Engineering, University of Sydney, Australia, and with the Graduate School of Information Sciences, Tohoku University, Japan (e-mail: a.jamalipour@ieee.org).}}
        \par \textit{(Corresponding authors: Geng Sun and Jiahui Li.)}}

}

\markboth{Journal of \LaTeX\ Class Files,~Vol.~X, No.~X, June~2026}%
{Shell \MakeLowercase{\textit{et al.}}: Bare Demo of IEEEtran.cls for Computer Society Journals}


\IEEEtitleabstractindextext{%
\begin{abstract}		
An integration of satellites and terrestrial networks is crucial for enhancing performance of next-generation communication systems. However, the networks are hindered by the long-distance path loss and security risks in urban canyons. In this work, we propose a satellite-terrestrial covert communication system assisted by the aerial active transmissive reconfigurable intelligent surface (AAT-RIS) to improve the channel capacity while ensuring the transmission covertness. Specifically, we first derive the minimal detection error probability (DEP) under the worst condition that the Warden has perfect channel state information. Then, we formulate an AAT-RIS-assisted satellite-terrestrial covert communication optimization problem (ASCCOP) to maximize the sum of the fair channel capacity for all ground users while meeting the strict covert constraint, by jointly optimizing the trajectory and active beamforming of the AAT-RIS. Due to the challenges posed by the complex and high-dimensional state-action spaces as well as the need for efficient exploration in dynamic environments, we propose a generative deterministic policy gradient (GDPG) algorithm, which is a generative deep reinforcement learning-based method to solve the online ASCCOP. Concretely, the generative diffusion model is utilized as the policy representation of the proposed algorithm to enhance the exploration process by generating diverse and high-quality samples through a series of denoising steps. Moreover, we incorporate an action gradient mechanism to accomplish the policy improvement of the proposed algorithm, which refines the better state-action pairs through the gradient ascent. Simulation results demonstrate that the proposed approach significantly outperforms important benchmarks, and also validate the robustness under different algorithm parameters and environment settings.
\end{abstract}
	
}

\maketitle
\IEEEdisplaynontitleabstractindextext
\IEEEpeerreviewmaketitle

%
%
\section{Introduction}
\label{sec_introduction}

\par \IEEEPARstart{A}{s} a crucial component of the sixth-generation (6G) wireless communication systems, satellite-terrestrial communications have garnered significant attention in both military and civil sectors due to their extensive coverage \cite{Mahboob2024}. For instance, in the aftermath of natural disasters such as earthquakes and floods, satellite-terrestrial systems can be an indispensable solution to re-establish communication links with disabled ground-based facilities. However, the persistent challenges of high path loss from the long-distance transmission and extreme penetration loss caused by obstacles continue to affect the performance of these systems, especially in some urban canyon environments \cite{AlHourani2020}. 

\par In such cases, low-altitude platforms (LAPs) can be deployed as the aerial relay nodes between satellites and ground users \cite{Jiahui2025}. Compared to the fixed relay communication mode, LAP-based aerial relay communications can bring an additional degree of freedom by dynamically adjusting their trajectories, thereby improving the channel capacity for the users in various locations {\color{black}{\cite{Hu2021}}}. While the high line-of-sight (LoS) probability of air-to-ground channels improves the transmission performance, the inherent broadcast nature of wireless communications simultaneously heightens the security vulnerabilities \cite{Ding2024}. As such, two widely used conventional techniques for ensuring secure communications are the encryption method in the high-layer network protocol stack and physical layer security \cite{li2024two}. However, these methods cannot fundamentally address security threats of satellite-terrestrial communications, particularly in the scenarios involving advanced technologies such as quantum computers, which could potentially break the conventional encryption methods by adapting to exploit emerging decoding techniques \cite{Chen2023}. 

\par Given the limitations of conventional security methods in addressing new forms of threats, covert communications have been recognized as a promising alternative \cite{Kang2025}. Specifically, covert communications exploit the uncertainty of the Warden to prevent it from detecting the transmission by hiding the transmission information in noise, thereby providing a more fundamental communication security. This approach aligns well with the dynamic capabilities of reconfigurable intelligent surfaces (RIS) and their variants, which adjust wireless channels by controlling intelligent reflection elements, thus positioning them as crucial technologies in covert communications for terrestrial environments \cite{WangMan2025}. Among the available RIS solutions, the active transmissive RIS can become the preferred choice for mounting on LAPs in the satellite-terrestrial covert communication systems \cite{SongR2024}. This is because an active transmissive RIS mounted horizontally on a LAP can establish a direct satellite-terrestrial link by transmitting incident signals from the upper to the lower hemisphere, while also amplifying the signal to maintain a high transmission rate under diverse and dynamic channel conditions. Therefore, this paper aims to explore the satellite-terrestrial covert communications with the aid of the aerial active transmissive RIS (AAT-RIS).

\subsection{Prior Works}
\label{sub_sec_prior_works}

{\color{black}{\par We review related works from three aspects, including the architecture of satellite-terrestrial covert communications, uncertainty sources and performance metrics in satellite-terrestrial covert communications, and optimization methods for covert communications, which motivates this work.}}

\subsubsection{Architecture of Satellite-Terrestrial Covert Communications}
\label{sub_sub_sec_Architecture of Satellite-Terrestrial Covert Communications}

\par Existing works in the architecture of satellite-terrestrial covert communications can be mainly divided into direct mode and relay mode. For the direct mode, Zhang \textit{et al}. \cite{Zhanglei2024} investigated an ultra-dense low Earth orbit (LEO) satellite downlink system, where the sidelobe leakages of other satellites are utilized to ensure the covertness. In \cite{Jia2025}, the authors developed a direct satellite-terrestrial downlink system by employing the rate-splitting multiple access technique to transmit jamming signals. However, this mode typically overlooks the long-distance path loss and complex propagation environments between satellites and ground users, which are particularly problematic in dense urban areas. In terms of the relay mode, Wu \textit{et al}. \cite{Wu2022} studied the ground device-based satellite-terrestrial covert relay communications, where multiple ground base stations are randomly selected to relay the wireless signal. {\color{black}{Moreover, the authors in \cite{Song2023} presented a RIS-based geosynchronous earth orbit (GEO) satellite-terrestrial covert relay communication system, where a passive RIS is deployed on the surface of the building to optimize the signal reflection.}} However, the fixed position of relay platforms limits their adaptability in dynamic environments and reduces the flexibility of the system.

\subsubsection{Uncertainty Source and Performance Metrics in Satellite-Terrestrial Covert Communications}
\label{sub_sub_sec_Uncertainty Source and Performance Metrics in Satellite-Terrestrial Covert Communications}

\par In the field of satellite-terrestrial covert communications, several studies have explored modeling the uncertainty for the Warden from different sources and designing the corresponding performance metrics. For example, Guo \textit{et al}. \cite{Guo2024} presented a full-duplex ground receiver-based architecture of the satellite-terrestrial covert communications, where the transmit antenna of the ground receiver is used to send the jamming signal to raise the uncertainty of the Warden. In \cite{Yu2024}, the authors investigated the public message as a cover to produce the received power uncertainty in satellite-terrestrial covert communications, where the covert rate is maximized by jointly optimizing the overt message power allocation factor. Moreover, Feng \textit{et al}. \cite{Feng2024} proposed the passive RIS-assisted multi-satellite cooperative covert systems and utilized the active beamforming of multiple satellites and passive beamforming of RIS to achieve the communication covertness, while ensuring the maximization of the received signal power. However, these works typically focus on introducing uncertainty through additional components such as dedicated jamming devices, which overlook the inherent noise uncertainty of the environment itself. Moreover, the majority of existing works primarily center on optimizing the covert communication rate with limited consideration given to the fairness between users.
{\color{black}{
\subsubsection{Optimization Methods for Covert Communications}
\label{sub_sub_sec_Optimization Methods for Satellite-Terrestrial Covert Communications}
}}
\par Existing studies on optimization methods for covert communications primarily relied on convex optimization theory. For example, Hui \textit{et al}. \cite{Hui2024} proposed a two-stage iterative method with the block coordinate descent to maximize the average detection error probability (DEP). In \cite{Cao2024}, the authors utilized convex optimization to minimize the average age-of-information by taking the derivative of the transmission power of the transmitter in satellite-terrestrial covert communications. Moreover, Jia \textit{et al}. \cite{Jia2025} maximized the minimum covert rate by using the fractional programming, semidefinite relaxation and the S-procedure. {\color{black}{Furthermore, Xing \textit{et al}. \cite{Xing2024} proposed a covert transmission scheme for cell-free Internet-of-Thing networks that leverages the Lagrangian dual algorithm and semidefinite relaxation to jointly optimize beamforming and artificial noise vectors, thereby improving the covert transmit rate.}} However, these methods often assume static environments or require the channel state information (CSI) of the Warden. These assumptions limit the practicality of these methods in satellite-terrestrial covert communication systems, where the instantaneous CSI of the Warden is difficult to obtain and the detection threshold of the Warden may evolve over time.

\subsection{Motivations and Contributions}
\label{sub_sec_Motivations and Contributions}

\par As discussed above, existing satellite-terrestrial covert communication systems still face significant challenges in terms of the system design and optimization. To overcome these challenges, the new solution can be designed with three critical considerations. \textit{From the perspective of the system architecture}, a flexible relay system is needed that enhances both the degree of freedom and possesses the capability of amplifying signals to improve the transmission rate of the system. \textit{From the perspective of uncertainty sources and performance metrics}, the fairness among users must be incorporated to ensure that all users experience equitable performance while maintaining covertness without introducing extra components. \textit{From the perspective of the optimization methods}, an efficient optimization method should be able to adapt to dynamic environments online without relying on excessive prior knowledge, thereby enabling effective decision-making even under uncertain environment conditions.

\par Accordingly, this paper explores an AAT-RIS-assisted satellite-terrestrial covert communications and presents a generative deep reinforcement learning (DRL) algorithm to optimize the sum of the fair channel capacity for all ground users in dynamic environments. The main contributions of this paper are listed as follows: 
\begin{itemize}
\item \textit{AAT-RIS-assisted Satellite-Terrestrial Covert Communication Systems}: We propose a novel AAT-RIS-assisted satellite-terrestrial covert communication system that leverages an LAP equipped with active transmissive RIS to act as a relay between a GEO satellite and the ground users. This is particularly beneficial in urban canyon scenarios where the direct link is limited. Specifically, the system utilizes the flexibility of the LAP and the reconfigurable capability of an active transmissive RIS to dynamically adjust wireless channel properties, thereby enhancing the transmission rate by the signal amplification and concentration while ensuring the covert transmission by manipulating the electromagnetic wave propagation.

\item \textit{Covert Requirement Derivation and Optimization Problem Formulation}: Under the worst-case condition that the Warden has the perfect CSI information, we derive the covert requirement of the considered system based on the uncertainty of environmental noise. To further enhance the transmission rate and ensure the fairness among the ground users in such cases, we formulate an AAT-RIS-assisted satellite-terrestrial covert communication optimization problem (ASCCOP) by jointly optimizing the trajectory and active beamforming of the AAT-RIS. Notably, the formulated ASCCOP is proven as a non-convex long-term optimization problem.

\item \textit{Generative Online DRL Algorithm Design}: Since the formulated ASCCOP is a dynamic long-term optimization problem and involves the strict covert constraint, we propose a generative deterministic
policy gradient (GDPG) algorithm to efficiently explore the decision space. Specifically, the proposed GDPG algorithm integrates the generative diffusion model (GDM) into the deterministic policy gradient (DPG) algorithm as the multimodal policy representation in the complex and high-dimensional state-action spaces, thereby generating diverse and high-quality samples through a series of denoising steps to explore the state-action space more efficiently. Moreover, we incorporate an action gradient mechanism to guide the policy towards the higher-value regions by refining better state-action pairs with the gradient ascent method. 

\item \textit{Performance Validation}: Simulation results demonstrate that the proposed approach outperforms important benchmarks in terms of the sum channel capacity of all ground users, fairness among all ground users, and covertness of the system. 
\end{itemize}

\subsection{Organization}
\label{sub_sec_Organization and Notations}

\par The remainder of this paper is organized as follows. We first describe the system model in Section \ref{sec_system_model}. In Section \ref{sec_problem_formulation_and_Analysis}, the covert requirement of the system is derived and ASCCOP is formulated and analyzed. Next, we present a generative DRL algorithm to solve the proposed ASCCOP in Section \ref{sec_proposed_algorithm}. Section \ref{sec_performance_evaluation} shows numerical results and provides corresponding discussions. Finally, Section \ref{sec_conclusion} concludes this paper.


%
%
\section{System Model}
\label{sec_system_model}

\par In this section, we first describe the considered satellite-terrestrial communication scenario. Then, the AAT-RIS, channel, and communication models are given in detail. Finally, the binary hypothesis testing at the Warden is presented.
\subsection{Scenario Description}
\label{sub_sec_scenario_description}

\begin{figure}[t]
	\centering
	\includegraphics[width=\linewidth]{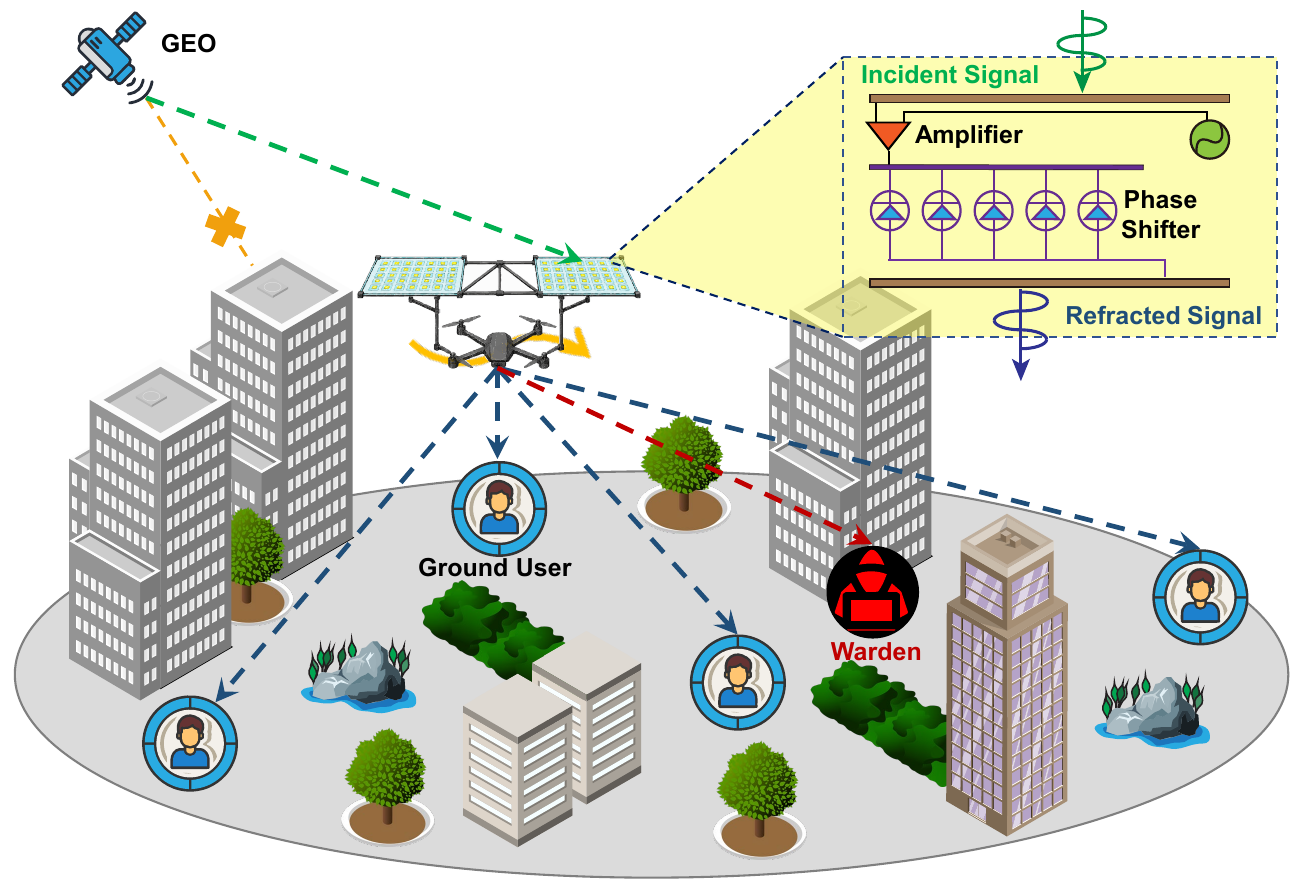}
	\caption{An illustration of the covert satellite-terrestrial downlink communication system, where an AAT-RIS acts as a relay, thereby enabling signal amplification and reconfiguration to forward covert information from the GEO satellite to multiple ground users without detection by a Warden.}
	\label{Fig: System Model}
\end{figure}

\par As shown in Fig. \ref{Fig: System Model}, we consider that a covert satellite-terrestrial downlink communication system, where a GEO satellite $a$ intends to communicate covertly with multiple ground users $\mathscr{K} \triangleq \{1, \dots, K\}$ against a passive ground Warden $w$. Due to the long distance and dense urban obstructions, we assume that the direct communication links between the GEO satellite and ground users are hindered by the low-gain receiving antennas and high propagation loss. To overcome this, the AAT-RIS{\color{color_revise}{\footnote{{\color{color_revise}{The AAT-RIS can be horizontally mounted on various low aerial platforms to ensure wide elevation-angle coverage and reduce aerodynamic drag during horizontal mobility, such as UAVs or electric vertical take-off and landing (eVTOL) aircraft, following the manner described in Ref. \cite{WDLukito2024}.}}}}} $r$ is deployed with $M = M_{x} \times M_{y}$ transmission and reflection elements, denoted by set $\mathscr{M} \triangleq \{1, \dots, M\}$, to enhance covert communication between the GEO satellite and ground users\footnote{In some emergency scenarios, GEO satellites can offer robust and continuous communication coverage, making them a broadly applicable solution. Our future work will consider incorporating LEO satellites, which may be suited to certain specific scenarios.}.

\par Without loss of generality, three-dimensional Cartesian coordinates are adopted to denote the positions of all pertinent communication nodes. In our considered scenario, all of the GEO satellite, ground users, and Warden are assumed to be stationary, with their coordinates denoted as $\mathbf{q}_{a} = \left[x_{a}, y_{a}, z_{a}\right]^{\mathrm{T}}$, $\mathbf{q}_{k} = \left[x_{k}, y_{k}, 0\right]^{\mathrm{T}}$, and $\mathbf{q}_{w} = \left[x_{w}, y_{w}, z_{w}\right]^{\mathrm{T}}$, respectively. Moreover, the total communication period $T$ is divided into $N$ time slots with an equal separation of $\delta_{t} = T/N$ and each time index is denoted as $n \in \mathscr{N} \triangleq \{1, \dots, N\}$. At time slot $n$, the position of the AAT-RIS is denoted as $\mathbf{q}_{r}[n] = \left[x_{r}[n], y_{r}[n], H\right]^{\mathrm{T}}$ with the fixed altitude $H$ to avoid collisions \cite{Wang2023}. To facilitate analysis and optimization, $\delta_{t}$ is chosen to be sufficiently small, thereby ensuring that the network environment remains quasi-static within each time slot. {\color{color_revise}{Moreover, for our considered scenario, we make the following assumptions about CSI as follows.

\begin{itemize}
    \item CSI available to the AAT-RIS: The AAT-RIS is assumed to have access to the instantaneous CSI from the GEO satellite and all ground users, which can be accomplished by well-established techniques in cooperative wireless communication systems \cite{Zheng2022, Guo2025}. However, since we model the Warden as a passive and non-cooperative adversary, the CSI from the AAT-RIS to the Warden is assumed to be entirely unknown to the AAT-RIS.
    \item CSI available to the Warden: To ensure rigorous covertness constraint modeling, we assume that the Warden possesses perfect instantaneous CSI for all relevant links, as adopted in \cite{Shahzad2018, He2025}.
\end{itemize}}}

\par In what follows, we introduce the AAT-RIS model, channel model and communication model to characterize the dynamics of the considered system and describe the relationship between the performance metrics and decision variables. 

\subsection{AAT-RIS Model}
\label{sub_sec_astar_ris_model}

\par Due to the horizontal mounting configuration of the transmissive RIS, the transmission coefficient matrix of the AAT-RIS at time slot $n$ can be given as follows \cite{Lizhen2024}:
\begin{equation}
	\label{Equ: Star_RIS_matrix}
	\begin{aligned}
		\mathbf{\Phi}[n] = \mathrm{diag}\Big(\Big[\beta_{1}[n]e^{j\phi_{1}[n]}, \dots, \beta_{M}[n]e^{j\phi_{M}[n]}\Big]\Big),
	\end{aligned}
\end{equation}
\noindent where $\beta_{m}[n]$ and $\phi_{m}[n]$ are the amplification gain and phase shift of the $m$-th element for the transmission space at time slot $n$, respectively.
\par To further enhance the transmission rate and fairness of the considered system, the AAT-RIS is designed to dynamically adjust its trajectory. Specifically, the position of the AAT-RIS at time slot $n$ can be calculated as follows:
\begin{equation}
	\label{Equ: AAT-RIS_position}
	\mathbf{q}_{r}[n] = \mathbf{q}_{r}[n-1] + \mathbf{v}_{r}[n-1] \cdot \delta_{t},
\end{equation}
\noindent where $\mathbf{v}_{r}[n-1] = \left[v_{x}[n-1], v_{y}[n-1], 0\right]^{\mathrm{T}}$ is the velocity of the AAT-RIS at time slot $n$.

\subsection{Channel Model}
\label{sub_sec_channel_model}

\par Let $\mathbf{h}_{ar}[n]$ and $\mathbf{h}_{rk}[n]$ stand for the channel gains from the GEO satellite to the AAT-RIS, from the AAT-RIS to the ground user $k$ and from the AAT-RIS to the Warden at time slot $n$, respectively. In this work, we consider that the channels from the GEO satellite to the AAT-RIS and from the AAT-RIS to the ground user $k$ and the Warden are Rician fading channels \cite{Zheng2025}, which can be formulated at time slot $n$ as follows \cite{Wang2025}:
\begin{equation}
	\label{Equ: Rician_channel}
\begin{aligned}
	\mathbf{h}_{\mathcal{l}}[n] = &\sqrt{L_{0} (d_{\mathcal{l}}[n])^{-\alpha_{\mathcal{l}}}} \Big(\sqrt{\frac{\kappa_{\mathcal{l}}}{\kappa_{\mathcal{l}}+1}}\mathbf{h}_{\mathcal{l}}^{\mathrm{LoS}}[n] \\&+ \sqrt{\frac{1}{\kappa_{\mathcal{l}}+1}}\mathbf{h}_{\mathcal{l}}^{\mathrm{NLoS}}[n]\Big),
	\mathcal{l} \in \{ar, rk, rw\},
\end{aligned}
\end{equation}
\noindent where $L_{0}$ represents the path loss at the reference distance of $1$ m, $d_{\mathcal{l}}[n]$ refers to the distance between the transmitter and receiver at time slot $n$, and $\kappa_{l}$ is the Rician factor. Moreover, $\mathbf{h}_{\mathcal{l}}^{\mathrm{LoS}}[n]$ and $\mathbf{h}_{\mathcal{l}}^{\mathrm{NLoS}}[n]$ are the LoS and the non-LoS (NLoS) components of the link $\mathcal{l}$, respectively. Specifically, the NLoS component can be modeled as the complex Gaussian distribution, i.e., $\mathbf{h}_{\mathcal{l}}^{\mathrm{NLoS}}[n] \sim \mathscr{CN}(\mathbf{0}, \mathbf{I})$. For the communication link from the GEO satellite to the AAT-RIS, the LoS component at time slot $n$ can be expressed as follows \cite{Lu2021}:
\begin{equation}
    \label{Equ: GEO_transmissive RIS_link_LoS}
\begin{aligned}
	\mathbf{h}_{ar}^{\mathrm{LoS}}[n] &= \left[1,\dots,e^{-j\frac{2\pi}{\lambda}\tilde{d}_{x}(M_{x}-1)\sin\varphi_{ar}[n]\cos\phi_{ar}[n]}\right]^{\mathrm{T}}\\&\otimes\left[1,\dots,e^{-j\frac{2\pi}{\lambda}\tilde{d}_{y}(M_{y}-1)\sin\varphi_{ar}[n]\sin\phi_{ar}[n]}\right]^{\mathrm{T}},
\end{aligned}
\end{equation}
\noindent where $\tilde{d}_{x}$ and $\tilde{d}_{y}$ represent the antenna spacing along the $x$-dimension and $y$-dimension, respectively. Moreover, $\sin\varphi_{ar}[n]\cos\phi_{ar}[n]$ and $\sin\varphi_{ar}[n]\sin\phi_{ar}[n]$ are the spatial frequency along the $x$-dimension and $y$-dimension corresponding to the angle of arrival $\varphi_{ar}[n]$ and $\phi_{ar}[n]$, respectively. Furthermore, $\sin\varphi_{ar}[n]\cos\phi_{ar}[n]$ and $\sin\varphi_{ar}[n]\sin\phi_{ar}[n]$ can be directly calculated by $(x_{r}[n]-x_{a}) / \Vert \mathbf{q}_{r}[n] - \mathbf{q}_{a} \Vert$ and $(y_{r}[n]-y_{a}) / \Vert \mathbf{q}_{r}[n] - \mathbf{q}_{a} \Vert$, respectively.

\par Similarly, the LoS component of the communication link from the AAT-RIS to the ground user $k$ and the Warden at time slot $n$ can be respectively expressed as follows:
\begin{equation}
	\label{Equ: transmissive RIS_User_link_LoS}
\begin{aligned}
    &\mathbf{h}_{\mathcal{l}}^{\mathrm{LoS}}[n] = \left[1,\dots,e^{-j\frac{2\pi}{\lambda}\tilde{d}_{x}(M_{x}-1)\sin\varphi_{\mathcal{l}}[n]\cos\phi_{\mathcal{l}}[n]}\right]^{\mathrm{T}}\\&\otimes\left[1,\dots,e^{-j\frac{2\pi}{\lambda}\tilde{d}_{y}(M_{y}-1)\sin\varphi_{\mathcal{l}}[n]\sin\phi_{\mathcal{l}}[n]}\right]^{\mathrm{T}}, \mathcal{l}\in\{rk, rw\},
\end{aligned}
\end{equation}
\noindent where $\sin\varphi_{\mathcal{l}}[n]\cos\phi_{\mathcal{l}}[n]$ and $\sin\varphi_{\mathcal{l}}[n]\sin\phi_{\mathcal{l}}[n]$ are the spatial frequency along the $x$-dimension and $y$-dimension corresponding to the angle of departure $\varphi_{\mathcal{l}}[n]$ and $\phi_{\mathcal{l}}[n]$, respectively.

\subsection{Communication Model}
\label{sub_sec_communication_model}

\par In the AAT-RIS-assisted satellite-terrestrial downlink communication system, the received signal at the ground user $k$ for the $i$-th channel at time slot $n$ can be represented as follows:
\begin{equation}
\label{Equ: received_signal_user}
\begin{aligned}
	\mathbf{y}_{k}^{i}[n] = \mathbf{h}_{rk}^{\mathrm{H}}[n]\mathbf{\Phi}[n]\left(\sqrt{p_{a}g_{a}}\mathbf{h}_{ar}[n] \mathbf{s}^{i}[n] + \mathbf{z}_{r}[n]\right) + z_{k}^{i}[n],
\end{aligned}
\end{equation}
\noindent where $p_{a}$ and $g_{a}$ represent the transmit power and transmit gain of the GEO satellite. Moreover, $z_{k}^{i}[n] \sim \mathscr{CN}(0, \sigma_{k}^{2})$ denotes the additive white Gaussian noise (AWGN) at the ground user $k$, and $\mathbf{z}_{r}[n] \sim \mathscr{CN}(\mathbf{0}, \sigma_{r}^{2}\mathbf{I})$ represents the AWGN introduced by the power amplifier at the AAT-RIS. 

\par Furthermore, the channel capacity between the GEO satellite and ground user $k$ at time slot $n$ can be expressed as follows:
\begin{equation}
	\label{Equ: channel_capacity_user}
	\begin{aligned}
		r_{ak}[n] = \log_{2}\Bigg(1 + \frac{p_{a}g_{a}\Big|\mathbf{h}_{rk}^{\mathrm{H}}[n]\mathbf{\Phi}[n]\mathbf{h}_{ar}[n]\Big|^{2}}{\sigma_{r}^{2}\Big\Vert\mathbf{h}_{rk}^{\mathrm{H}}[n]\mathbf{\Phi}[n]\Big\Vert^{2} + \sigma_{k}^{2}}\Bigg).
	\end{aligned}
\end{equation}

\subsection{Binary Hypothesis Testing at Warden}
\label{sub_sec_binary_hypothesis_testing_at_Warden}

\par Let the null hypothesis $\mathcal{H}_{0}$ and alternative hypothesis $\mathcal{H}_{1}$ denote the two circumstances that the GEO satellite is silent and sends the covert message to ground users, respectively. In particular, the Warden determines whether the GEO satellite is transmitting on the basis of the power of the signal received by a radiometer \cite{Chen2023}. For the $i$-th channel at time slot $n$, the received signal at the Warden from the GEO satellite can be described as follows:
\begin{equation}
	\label{Equ: binary_hypothesis_testing}
	\mathbf{y}_{w}^{i}[n]=
    \begin{cases}
    	\begin{aligned}
    		&\mathbf{h}_{rw}^{\mathrm{H}}[n]\mathbf{\Phi}[n]\left(\sqrt{p_{a}g_{a}}\mathbf{h}_{ar}[n] \mathbf{s}^{i}[n] + \mathbf{z}_{r}[n]\right)\\
    		& + z_{w}^{i}[n],
    	\end{aligned} &  \mathcal{H}_{1},\\
    	\begin{aligned}
    		& z_{w}^{i}[n],
    	\end{aligned} & \mathcal{H}_{0}, \\
	\end{cases}
\end{equation}
\noindent where $z_{w}^{i}[n] \sim \mathscr{CN}(0, \sigma_{w}^{2})$ are the AWGN at the Warden.

\par In the binary testing problem, the decision rule of the Warden can be expressed as follows {\color{black}{\cite{Hieu2023}}}:
\begin{equation}
	\label{Equ: decision_rule}
\begin{aligned}
	T_{w}[n] = \frac{1}{L}\sum_{i=1}^{L}\left|y_{w}^{i}[n]\right|^{2} \mathop{\gtrless}_{\mathcal{D}_{0}}^{\mathcal{D}_{1}}\tau[n],
\end{aligned}
\end{equation}
\noindent where $L$ represents the number of signal samples of the Warden, and $\tau[n]$ represents the decision threshold of the Warden at time slot $n$. Moreover, $\mathcal{D}_{0}$ and $\mathcal{D}_{1}$ are the decision results while supporting $\mathcal{H}_{0}$ and $\mathcal{H}_{1}$, respectively. Similar to \cite{Zhang2024}, we consider $L\rightarrow \infty$, which means that an infinite amount of samples can be used at the Warden.

\par As such, we can observe that optimizing the trajectory and active beamforming of the AAT-RIS can not only improve the channel capacity of ground users, but also reduce the risk of detection by the Warden. 

%
%
\section{Problem Formulation and Analysis}
\label{sec_problem_formulation_and_Analysis}

\par In this section, we first derive the covert requirement of the considered covert communication system at each time slot. Based on this, the ASCCOP is formulated and analyzed.

\subsection{Covert Requirement}
\label{sub_sec_covert_requirement}

\par The detection decision of the Warden involves two types of errors, which are the false alarm (FA) and miss detection (MD). The FA occurs when the Warden incorrectly assumes that the transmission is occurring while the GEO satellite remains silent. Conversely, the MD happens when the Warden fails to recognize that the GEO satellite is transmitting. Mathematically, the FA and MD probabilities are denoted as $P_{\mathrm{FA}}[n] = \Pr(\mathcal{D}_{1}|\mathcal{H}_{0})[n]$ and $P_{\mathrm{MD}}[n] = \Pr(\mathcal{D}_{0}|\mathcal{H}_{1})[n]$, respectively. Therefore, the DEP can be defined as follows:
\begin{equation}
    \label{Equ: detection_error_probability}
\begin{aligned}
	\xi[n] &= P_{\mathrm{FA}}[n]  + P_{\mathrm{MD}}[n]  \\
	&=\Pr(\sigma_{w}^{2} \ge \tau[n]) + \Pr(\iota[n] + \sigma_{w}^{2} \le \tau[n]),
\end{aligned}
\end{equation}
\noindent where $\iota[n] = p_{a}g_{a}\Big|\mathbf{h}_{rw}^{\mathrm{H}}[n]\mathbf{\Phi}[n]\mathbf{h}_{ar}[n]\Big|^{2} + \sigma_{r}^{2}\Big\Vert\mathbf{h}_{rw}^{\mathrm{H}}[n]\mathbf{\Phi}[n] \Big\Vert^{2}$.

\par Due to variations of the wireless communication environment, the noise uncertainty of the Warden can be described as follows \cite{Zhang2024}:
\begin{equation}
    \label{Equ: noise_uncertainty}
	f_{\sigma_{w}^{2}}(x)=
    \begin{cases}
		\begin{aligned}
			&\frac{1}{2 \ln (\rho) x}
		\end{aligned} &  \text {if } \frac{1}{\rho} \hat{\sigma}_{w}^{2} \leq x \leq \rho \hat{\sigma}_{w}^{2},\\
		\begin{aligned} & 0,
		\end{aligned} & \text {otherwise}, \\
	\end{cases}
\end{equation}
\noindent where $\rho$ and $\hat{\sigma}_{w}^{2}$ represent the noise uncertainty and nominal noise power, respectively. Considering the noise uncertainty and the scenario where the Warden has the perfect CSI, the DEP in Eq. \eqref{Equ: detection_error_probability} can be further formulated as follows:
\begin{equation}
\label{Equ: detection_error_probability_with_noise}
\begin{aligned}
	\xi[n] = 1 - \int_{\max\Big(\tau[n] - \iota[n], \frac{\hat{\sigma}_{w}^{2}}{\rho}\Big)}^{\tau[n]} \frac{1}{2 \ln (\rho) x}dx.
\end{aligned}
\end{equation}

\par Then, the derivative of $\xi[n]$ with respect to $\tau[n]$ can be calculated as follows:
\begin{small}
\begin{equation}
\label{Equ: the_derivative_of_xi}
\frac{\partial \xi[n]}{\partial \tau[n]}= \begin{cases}
		\begin{aligned}
			&-\frac{1}{(2 \ln \rho) \tau[n]},
		\end{aligned} &  \tau[n] \leq \iota[n]+ \frac{\hat{\sigma}_{w}^{2}}{\rho},\\
		\begin{aligned}
			& \frac{1}{2 \ln \rho}\left(\frac{1}{\tau[n]-\iota[n]}-\frac{1}{\tau[n]}\right),
		\end{aligned} & \tau[n]>\iota[n] + \frac{\hat{\sigma}_{w}^{2}}{\rho}. \\
	\end{cases}
\end{equation}
\end{small}

\par When $\tau[n] \leq \iota[n]+ \frac{\hat{\sigma}{w}^{2}}{\rho}$, the derivative is negative, thereby implying that $\xi[n]$ decreases with respect to $\tau[n]$. When $\tau[n] > \iota[n] + \frac{\hat{\sigma}{w}^{2}}{\rho}$, the derivative is positive, and thus $\xi[n]$ increases with respect to $\tau[n]$. Moreover, the optimal detection threshold should be within $[\frac{\hat{\sigma}_{w}^{2}}{\rho}, \rho \hat{\sigma}_{w}^{2}]$. Thus, the optimal detection threshold for the Warden at time slot $n$ can be expressed as follows:
\begin{equation}
\label{Equ: optimal_detection_threshold}
	\tau^{*}[n] = \min\Big(\iota[n] + \frac{\hat{\sigma}_{w}^{2}}{\rho}, \rho \hat{\sigma}_{w}^{2}\Big).
\end{equation}
\par Substituting $\tau^{*}[n]$ into Eq. \eqref{Equ: detection_error_probability_with_noise}, the minimal DEP can be calculated as follows:
\begin{small}
\begin{equation}
    \label{Equ: minimal_DEP}
	\xi^{*}[n] = 
    \begin{cases}
		\begin{aligned}
			&1 - \frac{1}{(2 \ln \rho) x}\ln\left(1 + \frac{\rho\iota[n]}{\hat{\sigma}_{w}^{2}}\right),
		\end{aligned} & \text {if } \iota[n] \leq \rho \hat{\sigma}_{w}^{2} - \frac{\hat{\sigma}_{w}^{2}}{\rho},\\
		\begin{aligned} & 0,
		\end{aligned} & \text {otherwise}. \\
	\end{cases}
\end{equation}
\end{small}

\par The AAT-RIS-assisted satellite-terrestrial covert communication system aims to ensure that $\xi^{*}[n] \geq 1 - \varepsilon$, where $\varepsilon$ is a small constant to control the covert requirement. Combining Eq. \eqref{Equ: minimal_DEP}, the considered system covert requirement at time slot $n$ can be reformulated as follows:
\begin{equation}
    \label{Equ: covert_constraint}
	\iota[n] < \min\Bigg(\rho \hat{\sigma}_{w}^{2} - \frac{\hat{\sigma}_{w}^{2}}{\rho}, \frac{\left(\rho^{2\varepsilon} -1\right)\hat{\sigma}_{w}^{2}}{\rho}\Bigg).
\end{equation}

\subsection{Problem Formulation}
\label{sub_sec_problem_formulation}

\par As outlined earlier, we define the fairness index of the channel capacity according to Jain’s index to guarantee the fairness among all ground users, which can be represented as follows \cite{Sediq2013}:
\begin{equation}
    \label{Equ: Jain's index}
	J[n] = \frac{\left(\sum_{k=1}^{K}r_{ak}[n]\right)^{2}}{K\left(\sum_{k=1}^{K}r_{ak}[n]^{2}\right)}.
\end{equation}

\par We aim to maximize the sum of the fair channel capacity for all ground users under the worst situation subject to the covert constraint via a joint design of the trajectory and active beamforming of the AAT-RIS. Defining $\mathbf{v_{r}} \triangleq \{\mathbf{v}_{r}[n], \forall{n} \}$, $\boldsymbol{\beta} \triangleq \{\beta_{m}[n], \forall{m, n}\}$ and $\boldsymbol{\phi} \triangleq \{\phi_{m}[n], \forall{m, n}\}$,  the ASCCOP can be formulated as follows:
\begin{subequations}
    \label{optimization_problem}
\begin{align}
    &\mathop{\operatorname{max}}\limits_{ \{\mathbf{v_{r}}, \boldsymbol{\beta}, \boldsymbol{\phi}\}}  \quad\sum_{n=1}^{N} J[n]\sum_{k=1}^{K}r_{ak}[n] \label{optimization_problem_objective}  \\
	&\quad~~\textrm{s.t.}  \quad~~~ \eqref{Equ: covert_constraint}, \\
	&\quad~~~~~~  \quad~~~\tiny\Bigg\Vert\mathbf{q}_{r}[n+1] - \mathbf{q}_{r}[n]\Bigg\Vert \leq v_{\text{max}} \cdot \delta_{t}, \quad \forall n \in \mathscr{N}, \label{optimization_problem_constraint_2} \\
	&\quad~~~~~~  \quad~~~\beta_{m}[n] > 1, \quad \forall m \in  \mathscr{M}, n \in \mathscr{N}  \label{optimization_problem_constraint_3},\\
	&\quad~~~~~~  \quad~~~\phi_{m}[n] \in[0,2 \pi], \quad \forall m \in  \mathscr{M}, n \in \mathscr{N}  \label{optimization_problem_constraint_4},\\
	&\quad~~~~~~  \quad~~~\tiny\Bigg\Vert\mathbf{\Phi}[n]\mathbf{h}_{ar}[n]\Bigg\Vert^{2} + \sigma_{r}^{2}\Bigg\Vert\mathbf{\Phi}[n]\Bigg\Vert^{2} < P_{\max}^{\mathrm{active}}, \quad \forall n \in \mathscr{N} \label{optimization_problem_constraint_5},\\
    &\quad~~~~~~  \quad~~~ X_{\min} \leq x_{r}[n] \leq X_{\max}, \quad \forall n \in \mathscr{N} \label{optimization_problem_constraint_6},\\
    &\quad~~~~~~  \quad~~~ Y_{\min} \leq y_{r}[n] \leq Y_{\max}, \quad \forall n \in \mathscr{N} \label{optimization_problem_constraint_7},
\end{align} 
\end{subequations}
\noindent where the constraints \eqref{optimization_problem_constraint_2}, \eqref{optimization_problem_constraint_3} and \eqref{optimization_problem_constraint_4} confine the velocity, amplification gains and phase shifts of the AAT-RIS, respectively. Moreover, the constraint \eqref{optimization_problem_constraint_5} is to ensure that the transmit power budget of the AAT-RIS must not exceed the maximum restriction. In addition, the constraints \eqref{optimization_problem_constraint_6} and \eqref{optimization_problem_constraint_7}  represent the boundary constraints for the AAT-RIS.

{\color{color_revise}{
\subsection{Problem Analysis}
\label{sub_sec_problem_analysis}

\par The formulated ASCCOP presents some specific challenges, which are analyzed as follows.
\begin{itemize}
    \item \textit{Complex Trade-offs among Objectives and Constraints}: Sum channel capacity, fairness among ground users and covertness constraint have inherent trade-offs. Specifically, maximizing the sum channel capacity tends to prioritize users with favorable channel conditions, thereby compromising fairness among ground users. Meanwhile, enhancing the sum channel capacity often necessitates increasing the transmit power of the AAT-RIS, which elevates the probability of detection by the Warden. Furthermore, enforcing fairness for ground users in disadvantaged locations typically requires increasing transmission power or maneuvering the AAT-RIS closer to those users. However, such behaviors cause a simultaneous decrease in covertness performance when the Warden is near such users.
    \item \textit{Non-convexity and Long-term Optimization}: The ASCCOP is inherently non-convex since the optimization objective is non-convex with respect to $\mathbf{v}_{r}$, $\boldsymbol{\beta}$ and $\boldsymbol{\phi}$, and the feasible solution space is also not a convex set due to the existence of constraints \eqref{Equ: covert_constraint} and \eqref{optimization_problem_constraint_5}. Moreover, since the goal is to maximize the fair channel capacity over a long-term horizon, the optimization requires taking into account future states of the system.
    \item \textit{Dynamic Environment and Imperfect Information}: The environment in which the AAT-RIS operates is highly dynamic with fluctuating channel conditions. Moreover, another significant challenge arises from the lack of real-time CSI about the Warden.
\end{itemize}

\par Conventional methods, e.g., convex optimization and dynamic programming, typically struggle with this because they require knowledge of future states, which are often uncertain and not directly observable in real-time. In this case, DRL has shown promise in addressing dynamic and uncertain environments in wireless communications \cite{Li2024}. Therefore, we transform the ASCCOP in the context of DRL and design an effective DRL-based algorithm to solve it.}}


\section{Proposed Algorithm}
\label{sec_proposed_algorithm}	

\par In this section, we first reformulate the considered ASCCOP as a Markov decision process. Then, we introduce the basics of the conventional DPG algorithm. Next, we design the GDPG algorithm to solve the proposed ASCCOP. Finally, we analyze the complexity of the proposed GDPG algorithm.

\subsection{MDP Formulation}
\label{sub_sec_MDP_formulation}

\par The dynamic decision process in the proposed ASCCOP can be naturally modeled as an MDP, which can be described as a $5$-tuple $\Omega = <\mathcal{S}, \mathcal{A}, R, P, \gamma>$ \cite{Yimengjie2025}. At each time slot $n$, the agent receives the state $\mathbf{s}[n]$ from the environment and selects an action $\mathbf{a}[n] \in \mathcal{A}$ based on its policy $\pi(\mathbf{a}|\mathbf{s})$. Then, the next state is generated based on the transition probability $P(\mathbf{s'}|\mathbf{s}, \mathbf{a})$, and the agent receives the reward $r[n] = R(\mathbf{s}[n], \mathbf{a}[n])$ from the environment as a feedback. Moreover, $\gamma \in (0,1]$ is the discount factor that balances immediate and future rewards. Accordingly, the details of each element in the considered ASCCOP are explained as follows. 

\subsubsection{Action}
\label{sub_sub_sec_action}

\par According to the design of ASCCOP, the action at time slot $n$ contains the trajectory control and active beamforming of the AAT-RIS, which can be defined as follows:
\begin{equation}
    \label{Equ: action space}
    \mathbf{a}[n] = \left\{\mathbf{v}_{r}[n], \beta_{1}[n], \dots, \beta_{M}[n], \phi_{1}[n], \dots, \phi_{M}[n]\right\},
\end{equation}
\noindent {\color{color_revise}{where the dimension of the action space is $2 + 2M$.}}
\subsubsection{State}
\label{sub_sub_sec_state}

\par At each time slot, the state space represents the observable information relevant to the decision-making process. In the context of ASCCOP, the state space consists of the following three components:

\begin{itemize} 
\item \textit{Position Information}: The system considers all entities remain at a constant altitude. This information is represented as the horizontal coordinates of the AAT-RIS, positions of ground users, and position of the Warden. 
\item \textit{Instantaneous CSI Information}: This part contains real-time CSIs between the GEO satellite and AAT-RIS as well as between the AAT-RIS and ground users. Note that the CSIs are decomposed into their real and imaginary components as the input of neural networks. 
\item \textit{Historical Information}: Because the historical information can provide necessary information for making decisions in the current time slot, last action, channel capacities between the GEO satellite and all users, and the received reward are contained in this part. 
\end{itemize}

\par Thus, the state $s[n]$ at time slot $n$ can be written as follows:
\begin{equation}
\label{Equ: state space}
\begin{aligned}
    \mathbf{s}[n] = \{&x_{r}[n], y_{r}[n], x_{1}, \dots, x_{K}, y_{1}, \dots, y_{K}, \\
    &x_{w}, y_{w}, \mathbf{h}_{ar}[n], \mathbf{h}_{r1}[n], \dots, \mathbf{h}_{rK}[n],\mathbf{a}[n-1],\\
    &r_{a1}[n-1], \dots, r_{aK}[n-1] ,r[n-1]\},
\end{aligned}
\end{equation}
\noindent {\color{color_revise}{where the dimension of the state space is $2M(K+1)+3K+9$.}}

{\color{color_revise}{\subsubsection{Reward}
\label{sub_sub_sec_reward}

\par The optimization objective of the formulated ASCCOP aims to maximize the sum of fair channel capacity among the ground users while satisfying the system constraints. To achieve this, the reward at time slot $n$ is defined as follows:
\begin{equation}
    \label{Equ: reward}
    \begin{small}
    \begin{aligned}
        r[n] &= J[n] \sum_{k=1}^{K} r_{ak}[n] \\
        &\quad - r_{pc} \cdot \mathbb{I} \left\{ \iota[n] \geq \min\Bigg(\rho \hat{\sigma}_{w}^{2} - \frac{\hat{\sigma}_{w}^{2}}{\rho}, \frac{\left(\rho^{2\varepsilon} -1\right)\hat{\sigma}_{w}^{2}}{\rho}\Bigg) \right\}  \\
        &\quad - r_{pr} \cdot \mathbb{I} \left\{ \left\Vert \bm{\Phi}[n] \mathbf{h}_{ar}[n] \right\Vert^{2} + \sigma_r^2 \left\Vert \bm{\Phi}[n] \right\Vert^2 \geq P_{\max}^{\mathrm{active}} \right\} \\
        &\quad - r_{pp} \cdot \mathbb{I} \left\{ \mathbf{q}_r[n+1] \notin [X_{\min}, X_{\max}] \times [Y_{\min}, Y_{\max}] \right\},
    \end{aligned}
\end{small}
\end{equation}

\noindent where $r_{pc}$ imposes a fixed penalty when the covert constraint is violated as determined by the theoretical detection error bound, $r_{pr}$ penalizes any violation of the AAT-RIS instantaneous power budget, and $r_{pp}$ penalizes trajectory violations beyond the operating boundaries. Moreover, the indicator function $\mathbb{I}\{\cdot\}$ returns 1 when the specified condition is satisfied and 0 otherwise. Note that the continuous action space with non-convex feasibility regions makes explicit illegal action analytically intractable, which motivates us to adopt the penalty-based reward design method rather than action masking.}}

\subsection{Conventional Deterministic Policy Gradient Algorithm}
\label{sub_sec_Conventional Deterministic Policy Gradient Algorithm}

\par The DPG algorithm is a classic reinforcement learning method that combines both policy-based and value-based methods \cite{Silver2014}, which enables simultaneous action generation and evaluation, thus improving training efficiency and decision-making in complex environments. Specifically, the main idea of the DPG algorithm primarily relies on policy evaluation and policy improvement to optimize decision-making processes. Let $\{\mathbf{s}[n], \mathbf{a}[n], \mathbf{s}[n+1], r[n]\}_{n \geq 0} \sim \pi_{\theta}$ be a trajectory sampled from the parameterized policy $\pi_{\theta}$, then the DPG algorithm aims to find an optimal policy $\pi_{\theta}^*$ that maximizes the expected cumulative return, which can be expressed as follows:
\begin{equation}
    \label{Equ: goal of DRL}
    \pi_{\theta}^* = \arg\max_{\pi_{\theta}} \mathbb{E}_{\pi_{\theta}} \left[ \sum_{n=0}^{N} \gamma^{n} R(\mathbf{s}[n], \mathbf{a}[n]) \right].
\end{equation}

\subsubsection{Policy Evaluation}
\label{sub_sub_sec_policy_evaluation}

\par The core objective of policy evaluation is to compute the state-action value function $Q_{\phi}^{\pi_{\theta}}(\mathbf{s},\mathbf{a})$ for a given policy $\pi_{\theta}$ in an environment. This function represents the expected cumulative return when following $\pi_{\theta}$, starting from state $s$ and taking action $\mathbf{a}$. Specifically, the Bellman equation of $Q_{\phi}^{\pi_{\theta}}(\mathbf{s},\mathbf{a})$ is given by:
\begin{equation}
    \label{Equ: Bellman of Q}
    Q_{\phi}^{\pi_{\theta}}(\mathbf{s}, \mathbf{a}) = \mathbb{E}_{\mathbf{s'} \sim P(\mathbf{s'}|\mathbf{s},\mathbf{a})} \left[ R(\mathbf{s},\mathbf{a}) + \gamma \mathbb{E}_{\mathbf{a'} \sim \pi_{\theta}} Q_{\phi}^{\pi_{\theta}}(\mathbf{s'}, \mathbf{a'}) \right].
\end{equation}

\par To minimize the difference between the estimated action-value function and the Bellman target, the loss function is defined as follows:
\begin{equation}
    \label{Equ: critic loss of DRL}
    \mathcal{L}(\phi) = \mathbb{E} \left[ \left| Q_{\phi}^{\pi_{\theta}}(\mathbf{s}, \mathbf{a}) - \left( R(\mathbf{s}, \mathbf{a}) + \gamma \mathbb{E}_{\mathbf{a'} \sim \pi_{\theta}} Q_{\phi}^{\pi_{\theta}}(\mathbf{s'}, \mathbf{a'}) \right) \right|^2 \right].
\end{equation}

\subsubsection{Policy Improvement}
\label{sub_sub_sec_policy_improvement}

\par The policy gradient theorem provides a foundation for policy improvement and this process can be expressed as follows \cite{Silver2014}:
\begin{equation}
    \label{Equ: policy improvement of DRL}
    \nabla_{\theta} J(\pi_{\theta}) = \mathbb{E}_{(\mathbf{s}, \mathbf{a}) \sim \pi_{\theta}} \left[ \nabla_{\theta} \log \pi_{\theta} (\mathbf{a} | \mathbf{s}) Q_{\phi}^{\pi_{\theta}}(\mathbf{s}, \mathbf{a}) \right].
\end{equation}

\par By iterating between policy evaluation and policy improvement, the DPG algorithm can converge to an optimal policy $\pi_{\theta}^*$.

\subsection{Generative Deterministic Policy Gradient Algorithm}
\label{sub_sec_Generative Deterministic Policy Gradient Algorithm}

{\color{color_revise}{
\subsubsection{Motivation of Adopting GDM}
\par When addressing the proposed ASCCOP, the conventional DPG algorithm faces two primary challenges. First, this method inherently struggles to explore the complex, high-dimensional and multimodal state-action spaces due to its reliance on unimodal Gaussian distribution \cite{Wang2023a}. Second, the frequent penalties imposed by covert constraints further exacerbate this issue by discouraging the agent from effectively leveraging the rare positive reinforcement signals. 

\par To overcome these limitations, we utilize GDM as the policy backbone. Different from fully-connected deep neural networks adopted in the conventional DPG algorithm, GDM generates diverse and high-fidelity action samples through iterative denoising, thereby enabling effective exploration across multimodal regions of the state-action space \cite{Yang2023}. Compared with other generative frameworks, GDM avoids the issue of non-differentiable propagation that arises in generative adversarial networks. Moreover, it also overcomes the limitations of variational autoencoders (VAE), which suffer from unimodal latent priors and susceptibility to posterior collapse \cite{R2-3-1}.}}

\begin{figure}[t]
	\removelatexerror
	\begin{algorithm}[H]
		\raggedright
		\caption{Policy Representation Based on GDM}
		\label{Algorithm 1: Policy Representation Based on Conditional GDM}
		\LinesNumbered
        \KwIn{$T$, $\boldsymbol{\epsilon}_\theta(\cdot)$, $\mathbf{s}[n]$}
        \KwOut{$\mathbf{a}[n]$}
        Initialize $\mathbf{x}_T \sim \mathscr{N}(\mathbf{0}, \mathbf{I})$\;
        \For{$t \gets T$ \textbf{to} $1$}{
            $\hat{\boldsymbol{\epsilon}} \gets \boldsymbol{\epsilon}_{\theta}(\mathbf{x}_t, \mathbf{s}[n], t)$\;
            $\mathbf{z} \sim \mathscr{N}(\mathbf{0}, \mathbf{I})$ if $t > 1$ else $\mathbf{z}=\mathbf{0}$\;
            $\mathbf{x}_{t-1}
            \gets \frac{1}{\sqrt{1 - \nu_t}}
            \Bigl(\mathbf{x}_t 
            - \frac{\nu_t}{\sqrt{1-\bar{\alpha}_t}} \,\hat{\boldsymbol{\epsilon}}\Bigr)
            + \sqrt{\nu_t}\,\mathbf{z}$\;
        }
        \Return $\boldsymbol{a}[n] \gets \mathbf{x}_0$;
	\end{algorithm}
\end{figure}

\subsubsection{GDM for Policy Representation}
\label{sub_sub_sec_GDM for Policy Representation}

\par According to \cite{Yang2023}, the basic process of GDM includes the forward process and reverse process. Specifically, the forward process of GDM incrementally corrupts a data sample by adding Gaussian noise over a series of time steps, thereby transforming its distribution into one that approximates pure Gaussian noise. More formally, this forward procedure constitutes a Markov chain, which can be defined as follows:
\begin{equation}
    \label{Equ: forward process of GDM}
    q(\mathbf{x}_{1:T}|\mathbf{x}_{0}) = \sum_{t=1}^{T} q(\mathbf{x}_{t}|\mathbf{x}_{t-1}),
\end{equation}
\noindent where $T$ is the total number of diffusion steps, and $q(\mathbf{x}_{t}|\mathbf{x}_{t-1}) = \mathscr{N}(\mathbf{x}_{t}; \sqrt{1-\nu_{t}}\mathbf{x}_{t-1}, \nu_{t}\mathbf{I})$ represents the transition kernel of the forward process. Moreover, $\nu_{t} \in (0,1)$ is a diffusion weight to control the speed of diffusion. After applying a series of derivations, the joint transition over all time steps can be marginalized to yield a closed-form expression for the distribution of $\mathbf{x}_{t}$ conditioned on $\mathbf{x}_{0}$, which can be computed as follows \cite{Zhang2025}:
\begin{equation}
    \label{Equ: the relationship between xt and x0}
    q(\mathbf{x}_{t}| \mathbf{x}_{0}) = \mathscr{N}\Big( \sqrt{\bar{\alpha}_{t}}\,\mathbf{x}_{0},\, \big(1-\bar{\alpha}_{t}\big)\mathbf{I}\Big),
\end{equation}
\noindent where $\bar{\alpha}_{t} = \prod_{i=1}^{t} \left(1-\nu_{i}\right)$. Given the closed-form expression, we can sample any $\mathbf{x}_{t}$ without iterating through the entire Markov chain.

\par The reverse process in GDM is to recover the original data distribution from a noise-corrupted sample by progressively removing the added noise at each time step. Specifically, the reverse process begins by initializing the latent variable at time step $T$ with $\mathbf{x}_T \sim \mathscr{N}(\mathbf{0}, \mathbf{I})$. Since the statistical distribution $q(\mathbf{x}_{t-1}| \mathbf{x}_{t})$ is intractable, a parameterized model is used to estimate this distribution, which can be expressed as follows:
\begin{equation}
    \label{Equ: unconverted parameterized model of reverse process}
    p_{\theta}(\mathbf{x}_{t-1}|\mathbf{x}_{t}) = \mathscr{N}(\mu_{\theta}(\mathbf{x}_{t}, t), \Sigma_{\theta}(\mathbf{x}_{t}, t)),
\end{equation}
\noindent where $\mu_{\theta}(\mathbf{x}_{t}, t)$ and $\Sigma_{\theta}(\mathbf{x}_{t}, t)$ represent the mean and the covariance matrix of the predicting noise at time step $t$, respectively. To incorporate specific semantic attributes into the reverse process, the conditional GDM can provide additional information to guide each denoising step \cite{Dhariwal2021}. Specifically, the reverse transition chain can be further expressed as follows:
\begin{equation}
    \label{Equ: conditional GDM reverse process}
    \small q(\mathbf{x}_{t-1}|\mathbf{x}_{t}, \mathbf{g})=
        \mathscr{N}\Big(\frac{1}{\sqrt{1 -\nu_{t}}}\Bigl(\mathbf{x}_{t}-\frac{\nu_{t}}{\sqrt{1-\bar{\alpha}_{t}}}\,\boldsymbol{\epsilon}_{\theta}(\mathbf{x}_{t}, \mathbf{g}, t)\Bigr), \nu_{t}\mathbf{I}\Big),
\end{equation}
\noindent where $\mathbf{g}$ is the condition information of reverse process. Furthermore, the parametrized noise-prediction network $\boldsymbol{\epsilon}_{\theta}(\mathbf{x}_{t}, \mathbf{g}, t)$ is used to estimate the noise component of the current data \(\mathbf{x}_t\) at each subsequent time step $t$. Moreover, a weighted version of the evidence lower bound (ELBO) can be adopted to learn $\boldsymbol{\epsilon}_{\theta}(\mathbf{x}_{t}, \mathbf{g}, t)$, which can be written as follows:
\begin{equation}
    \label{Equ: ELBO loss}
    \mathcal{L}(\theta) = \mathbb{E}\Big[\big|\epsilon - \epsilon_{\theta}(\sqrt{\bar{\alpha}_{t}}\mathbf{x}_{0}+\sqrt{1 - \bar{\alpha}_{t}}\mathbf{z}, \mathbf{g}, t)\big|^{2}\Big],
\end{equation}
\noindent where $\epsilon$ is a random sample from $\mathscr{N}(\mathbf{0}, \mathbf{I})$. To introduce the GDM as the policy representation into the DPG algorithm, we regard the reverse diffusion process as a generative mechanism that produces actions conditioned on the current state, which is illustrated in the Algorithm \ref{Algorithm 1: Policy Representation Based on Conditional GDM}.

\begin{figure}[t]
    \centering
    \includegraphics[width=\linewidth]{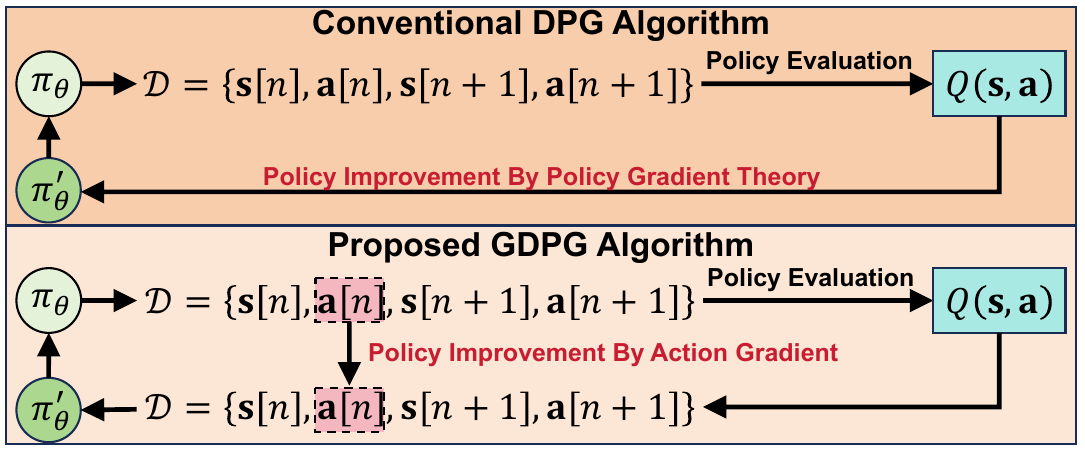}
    \caption{Comparison of the policy improvement between the conventional DPG algorithm and the proposed GDPG algorithm.}
    \label{Fig: Action Gradient}
\end{figure}

\begin{figure*}
    \centering
    \includegraphics[width=\linewidth]{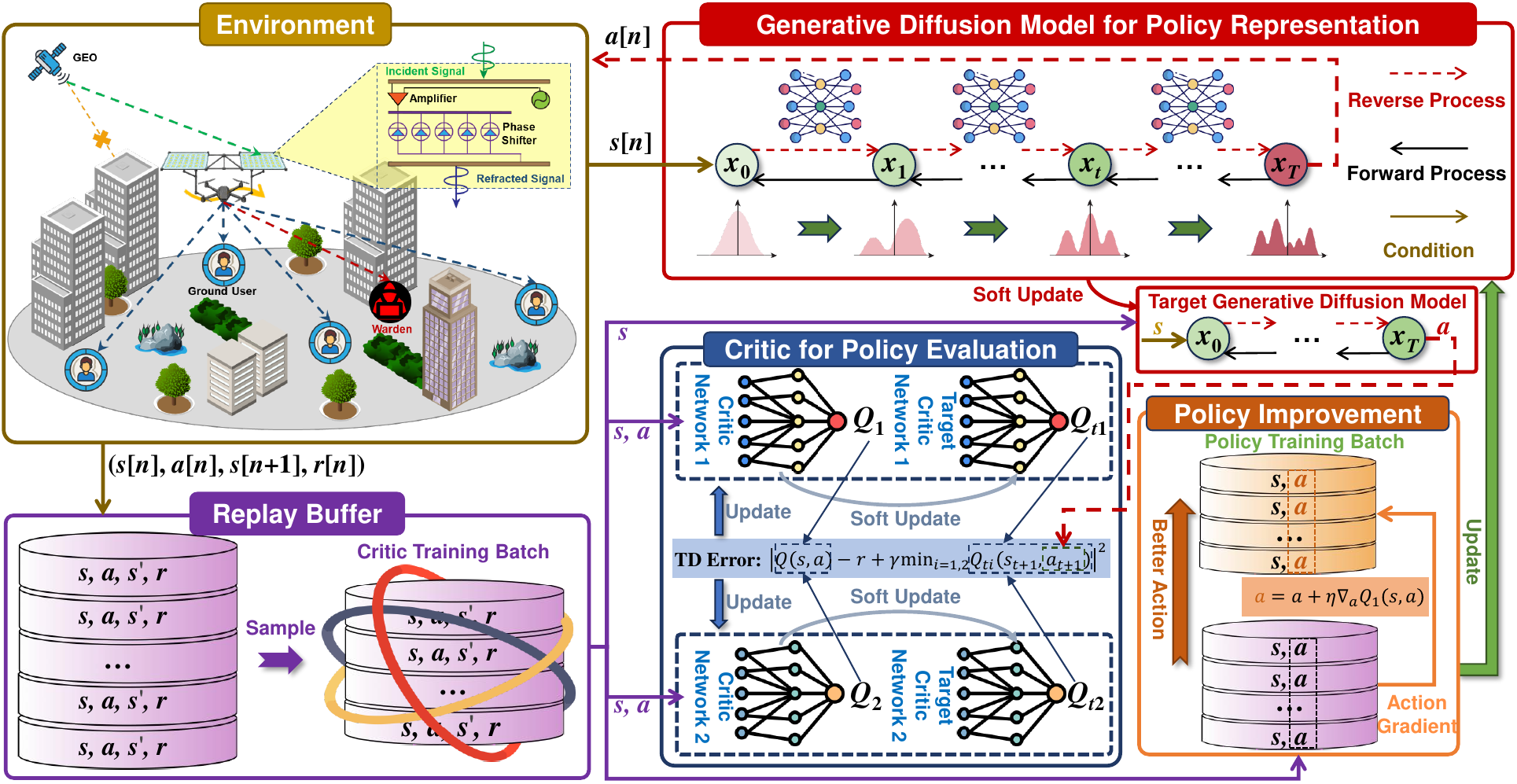}
    \caption{{\color{color_revise}{The overall architecture of the proposed GDPG algorithm to solve the ASCCOP, where GDM is used to execute the policy representation of the GDPG algorithm and action gradient mechanism is introduced to enable the policy improvement of the GDPG algorithm.}}}
    \label{Fig: Algorithm Framework}
\end{figure*}

\begin{figure}[t]
	\removelatexerror
	\begin{algorithm}[H]
		\raggedright
		\caption{GDPG Algorithm Training Procedure}
		\label{Algorithm 2: GDPGA Training Procedure}
		\LinesNumbered
        Initialize GDM network parameters $\theta$, critic network parameters $\phi_{1}$ and $\phi_{2}$ as well as their corresponding target network parameters ${t\theta}$, $t\phi_1$ and $t\phi_2$\;
        Initialize replay buffer $\mathcal{D}$\;
        \For{each episode}{
            Reset environment and initialize state $\mathbf{s}[0]$\;  
            \For{each time step $n$}{
                Select action $\mathbf{a}[n]$ by \textbf{Algorithm \ref{Algorithm 1: Policy Representation Based on Conditional GDM}}\;
                Execute action $\mathbf{a}[n]$, observe reward $r[n]$ and next state $\mathbf{s}[n+1]$\;
                Store transition $(\mathbf{s}[n], \mathbf{a}[n], r[n], \mathbf{s}[n+1])$ in replay buffer $\mathcal{D}$\;
                
                \If{current\_step $\geq$ start\_learning\_step}{
                    Sample a batch from $\mathcal{D}$\;
                    Compute the loss of critic networks by Eq. \eqref{Equ: critic-loss} and update critic networks by the gradient descent method\;              
                    Compute action gradients $\nabla_{\mathbf{a}} Q_{\phi}^{\pi_{\theta}}(\mathbf{s}, \mathbf{a})$ and update state-action pair of batch by Eq. \eqref{Equ: action gradient}\;
                    Compute weighted ELBO loss by Eq. \eqref{Equ: ELBO loss} and update GDM network\;
                    Soft update target networks by Eq. \eqref{Equ: soft update}
                }
            }
        }
	\end{algorithm}
\end{figure}

\subsubsection{Action Gradient for Policy Improvement}
\label{sub_sub_sec_Action Gradient for Policy Improvement}

\par As shown in Eq. \eqref{Equ: ELBO loss}, GDM merely learns to approximate the distribution of training samples produced by the current policy $\pi_{\theta}$. As such, it does not intrinsically offer a mechanism to optimize the policy beyond the scope of the data on which it is trained. Different from the conventional DPG algorithm, direct policy improvement through the policy gradient theorem is not feasible in this setting, as the diffusion-based policy is not parameterized straightforwardly but is instead learned via a stochastic process. Therefore, we adopt an action gradient mechanism to enable policy improvement, which is shown in Fig. \ref{Fig: Action Gradient}. Specifically, a given state-action pair $(\mathbf{s}, \mathbf{a})$ can be converted into a better state-action pair by the gradient ascent method, which can be represented as follows \cite{Zhang2025}:
\begin{equation}
    \label{Equ: action gradient}
    \mathbf{a} = \mathbf{a} + \eta_{a}\mathbb{E}_{(\mathbf{s},\mathbf{a})\in \mathcal{D}}\big[\nabla_{\mathbf{a}}Q_{\phi}^{\pi_{\theta}}(\mathbf{s}, \mathbf{a})\big]
\end{equation}
\noindent where $\eta_{a}$ is the step-size, and $\mathcal{D}$ represents the replay buffer used to store the experiences of the agent in the environment. Moreover, $\nabla_{\mathbf{a}}Q_{\phi}^{\pi_{\theta}}(\mathbf{s}, \mathbf{a})$ is referred to as the action gradient. By replacing the original state-action pair with the improved pair, the GDM-based policy can be updated continually, ensuring $\pi_{\theta}^{new} \succeq \pi_{\theta}$ and thereby accomplishing policy improvement.

\subsection{Overall Structure and Complexity Analysis}
\label{sub_sec_Overall Structure and Complexity Analysis}

\par Similar to approaches used in \cite{Fujimoto2018}, \cite{Haarnoja2018},  we employ a target GDM network $\pi_{t\theta}$, and a double critic networks $Q_{\phi_1}$ and $Q_{\phi_2}$ alongside their corresponding target critic networks $Q_{t\phi_1}$ and $Q_{t\phi_2}$ to avoid abrupt distribution shifts and address Q-value overestimation. Specifically, the loss function of the critic network can be expressed as follows:
\begin{equation}
    \label{Equ: critic-loss}
\begin{aligned}
    \mathcal{L}(\phi_i) = \mathbb{E}_{(\mathbf{s},\mathbf{a},r,\mathbf{s'}) \sim \mathcal{D}} \Big[\big|Q_{\phi_{i}}(\mathbf{s},\mathbf{a}) - y\big|^{2}\Big], \quad i \in \{1,2\},
\end{aligned}
\end{equation}
\noindent where $y = \Big(r + \gamma\min_{j \in \{1,2\}}Q_{t\phi_{j}}\!\big(\mathbf{s'},\pi_{t\theta}(\mathbf{s'})\big)\Big)$. Moreover, we periodically update the target networks for both the actor and the critic. Specifically, the parameters of the target networks are updated as follows:
\begin{equation}
    \label{Equ: soft update}
\begin{aligned}
    &t\theta \leftarrow \tau \theta + (1 - \tau) t\theta,\\
    &t\phi_i \leftarrow \tau \phi_i + (1 - \tau) t\phi_i, \quad i \in \{1,2\},
\end{aligned}
\end{equation}
where $\tau$ is a small positive constant that controls the rate of the soft update. 

\par The structure of the proposed GDPG algorithm for the ASCCOP is illustrated in Fig. \ref{Fig: Algorithm Framework} and the detailed implementation is outlined in Algorithm \ref{Algorithm 2: GDPGA Training Procedure}. Next, we analyze the computational and space complexity of the proposed GDPG algorithm for both training and execution phases. Specifically, the analysis focuses on the integration of GDM and DRL components, including diffusion-based action generation, dual-critic optimization, and action gradient mechanisms.

{\color{color_revise}{\subsubsection{Training Phase}
\label{sub_sub_sec_training_phase}

\par Let $G$, $B$, $|\boldsymbol{\phi}|$, $|\boldsymbol{\theta}|$ and $C$ represent the number of training episodes, batch size, the number of critic network parameters, the number of GDM network parameters, and the per-step environment interaction cost, respectively. Then, the total computational complexity during the training phase is $\mathcal{O}(GN(B+ C + (4+T)|\boldsymbol{\theta}| +6|\boldsymbol{\phi}|))$. Specifically, the computational complexity of the training phase is summarized as follows \cite{Zhang2025}:

\begin{itemize}
    \item \textit{Network Initialization}: This process involves initializing two critic networks, a GDM network, and their target networks, thereby increasing the complexity of $\mathcal{O}(4|\boldsymbol{\phi}| + 2|\boldsymbol{\theta}|)$.
    
    \item \textit{Action Selection and Environment Interaction}: GDM-based action generation at each step needs to performs $T$ denoising steps and interacts with the environment, thereby introducing the complexity of $\mathcal{O}(GN(T|\boldsymbol{\theta}| + C))$ during the entire training phase.
    
    \item \textit{Network Updates}: The overhead of the network update process includes several components. Specifically, the parameter update for the two critic networks incurs the complexity of $\mathcal{O}(2GN|\boldsymbol{\phi}|)$. Moreover, action gradient improvements add the complexity of $\mathcal{O}(GNB)$. Additionally, the GDM network update introduces the complexity of $\mathcal{O}(GNT|\boldsymbol{\theta}|)$. Finally, the soft update operation results in the complexity of $\mathcal{O}(GN|\boldsymbol{\phi}| + 2GN|\boldsymbol{\theta}|)$. As a result, the overall complexity is $\mathcal{O}(GN(B+ 2|\boldsymbol{\theta}| + 4|\boldsymbol{\phi}|))$.
\end{itemize}

\par Let $|\boldsymbol{s}|$ and $|\boldsymbol{a}|$ represent the dimensions of state and action spaces, respectively. Then, the space complexity during training phase is $\mathcal{O}(4|\boldsymbol{\phi}| + 2|\boldsymbol{\theta}| + D(2|\boldsymbol{s}| + |\boldsymbol{a}| + 1))$, which mainly includes the storage of network parameters $\mathcal{O}(4|\boldsymbol{\phi}| + 2|\boldsymbol{\theta}|)$ and the replay buffer storage $\mathcal{O}(D(2|\boldsymbol{s}| + |\boldsymbol{a}| + 1))$.}}

\subsubsection{Execution Phase}
\label{sub_sub_sec_execution_phase}

\par The computational complexity during execution is $\mathcal{O}(T|\boldsymbol{\theta}|)$, as each action decision requires $T$ denoising steps by the GDM network. The space complexity during the execution phase is $\mathcal{O}(|\boldsymbol{\theta}|)$, which involves only the storage of the GDM network parameters.

\begin{figure}
    \centering    \includegraphics[width=0.85\linewidth]{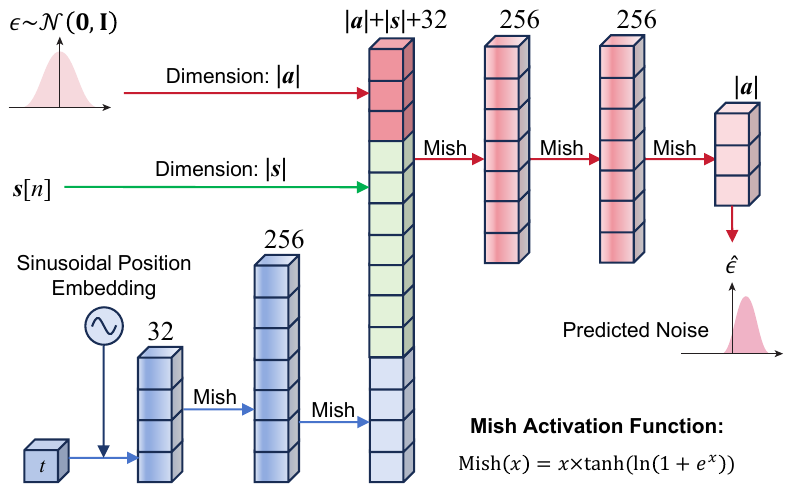}
    \caption{Architecture of GDM network with Mish activation function for non-monotonic gradient flow, and sinusoidal position embedding for efficient encoding of temporal information in the reverse process.}
    \label{Fig: Model Design}
\vspace{-5mm}
\end{figure}
\section{Performance Evaluation}
\label{sec_performance_evaluation}

\par In this section, we first describe our simulation setup. Then, comprehensive results are presented to assess the effectiveness and robustness of the proposed approach.

\subsection{Simulation Setup}
\label{sub_sec_simulation_setup}

\subsubsection{Environment, Platform and Model Details}
\label{sub_sub_sec_environment_and_model_details}

{\color{color_revise}{\par In this work, we consider the service area as a square region with 200 m by 200 m. In this region, 5 ground users are served by an LAP, operating at $H = 100$ m, and equipped with a $4 \times 4$ active transmissive RIS array to facilitate the wireless covert transmission between a GEO satellite and the ground users. Moreover, the simulations are conducted on a Linux server with an NVIDIA RTX 4090 GPU, using PyTorch 1.13, and CUDA 11.6. For the proposed GDPG algorithm, we utilize the Adam optimizer with a learning rate $3 \times 10^{-4}$ for both the GDM network and the critic networks, while setting $\eta_{a}=3 \times 10^{-2}$ for the action gradient. The specific GDM network structure is exhibited in Fig. \ref{Fig: Model Design} and other key parameters for the environment and algorithm are listed in Table \ref{Tab: Simulation Parameters}.}}
\begin{table}[t]
\centering
\renewcommand{\arraystretch}{1.2}
\caption{Simulation Parameters}
\label{Tab: Simulation Parameters}
\begin{tabular}{c|ll|ll}
\toprule[1.5pt] 
\multicolumn{1}{l}{\textbf{}}         & Parameter & Value & Parameter & Value \\ 
\hline
\multirow{8}{*}{\rotatebox{90}{Environment}} 
                     & $N$                                                      & 50            & $\delta_{t}$                      & 1 s       \\
                     & $\alpha_{ar}$                                            & 2.1           & $\alpha_{rk}$                     & 2.5       \\
                     & $\alpha_{rw}$                                            & 2.7           & $\kappa_{ar}$                     & 5         \\
                     & $\kappa_{kr}$                                            & 3             & $\kappa_{rw}$                     & 3         \\
                     & $v_{max}$                                                & 20 m/s        & $\lambda$                         & 0.15 m    \\
                     & $L_{0}$                                                  & -38 dB        & $\tilde{d}_x$, $\tilde{d}_y$      &  0.075 m  \\ 
                     & $p_a$                                                    & 59 dBW/MHz    & $g_a$                             & 51 dBi    \\ 
                     & $\sigma_{r}^{2}$, $\sigma_{k}^{2}$, $\sigma_{w}^{2}$     & -174 dBm/Hz  & $\rho$                            & 3 dB       \\ 
                     \hline
\multirow{3}{*}{\rotatebox{90}{Algorithm}} 
                     & $\gamma$             & 0.95              & $T$          &   10        \\
                     & $\tau$               &  0.005            & $B$          &   256       \\
                     & $D$           &  100000             & $G$          & 20000           \\
                     \bottomrule[1.5pt] 
\end{tabular}
\end{table}

\subsubsection{Benchmarks}
\label{sub_sub_sec_benchmarks}

\par To evaluate the effectiveness of the proposed approach and the GDPG algorithm, we compare both the overall approach and algorithm performances against several other approaches and algorithms.

\begin{itemize}
    \item \textit{Random}: At each time slot, actions are randomly selected from the action space. It is used primarily as a lower-bound baseline, which indicates the worst performance of the considered system.
    \item \textit{Non-active transmissive RIS (NA)}: In this approach, the transmissive RIS mounted on the LAP is passive, which does not amplify the transmission signals.
    \item \textit{Only Trajectory Optimization (OTO)}: This mode focuses solely on optimizing the trajectory of the AAT-RIS while keeping the active beamforming of the AAT-RIS fixed.
    \item \textit{Only Active Beamforming Optimization (OABO)}: In this case, only the active beamforming of the AAT-RIS is optimized, while the position of the AAT-RIS remains unchanged.
    {\color{color_revise}{\item \textit{State-of-the-Art DRL Algorithms}: The deep deterministic policy gradient (DDPG) and twin delayed DDPG (TD3) are the earliest actor-critic algorithms designed for continuous action spaces \cite{Fujimoto2018}. Moreover, soft actor-critic (SAC) is based on the maximum entropy reinforcement learning framework and also serves as a baseline \cite{Haarnoja2018}. Besides, we implement a VAE-enabled DPG algorithm by replacing the GDM with the VAE to enable direct assessment of GDM to multimodal exploration and policy stability.}}
\end{itemize}

\begin{figure}[t]
	\centering
	\includegraphics[width=\linewidth]{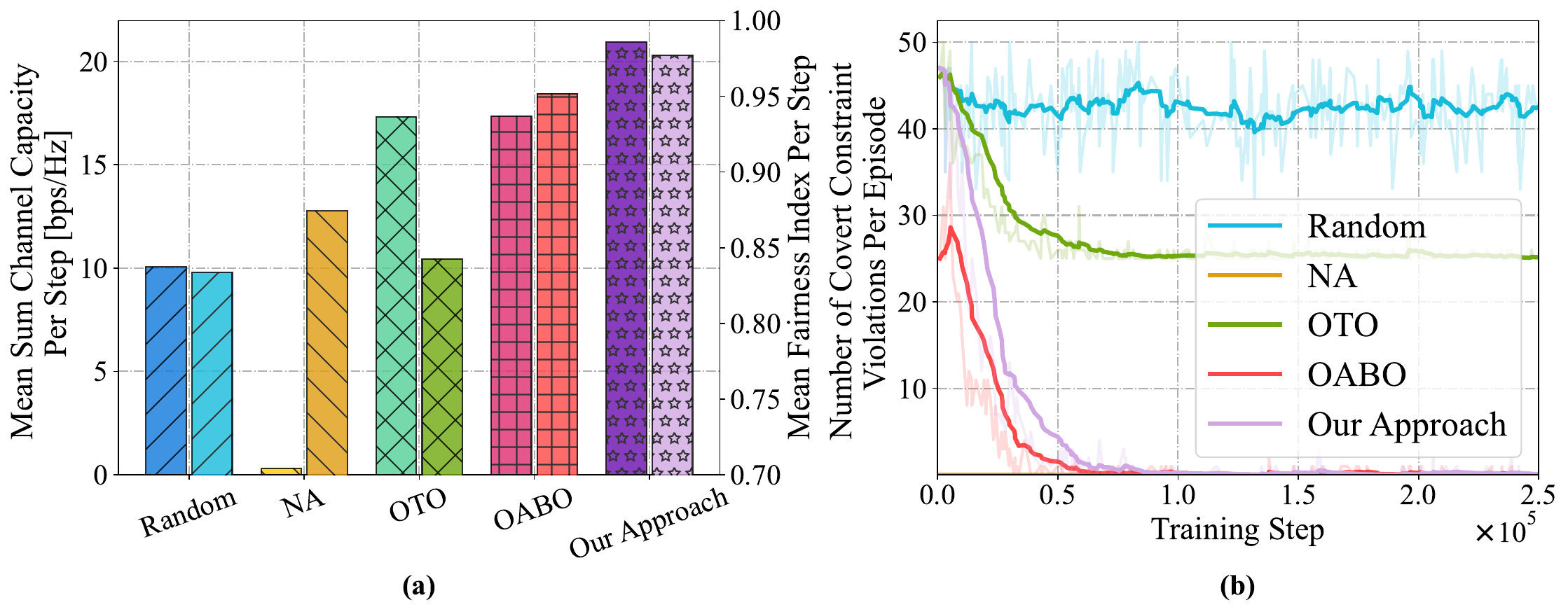}
	\caption{Performance comparison on different approaches. (a) Mean sum channel capacity and mean fairness index per step. (b) Number of covert constraint violations per episode.}
	\label{Fig: Different Approaches}
\vspace{-5mm}
\end{figure}

\begin{figure*}[htbp]
    \centering
    \includegraphics[width=\linewidth]{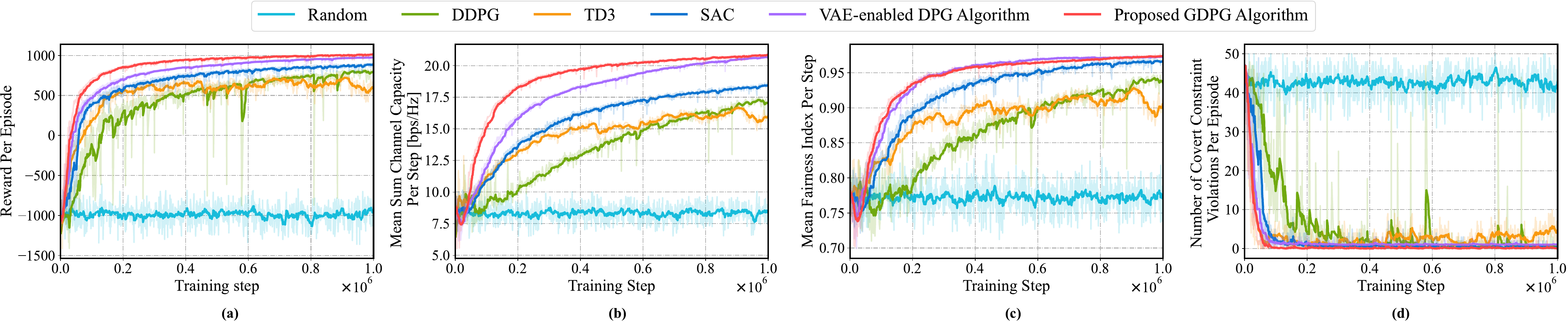}
    \caption{{\color{color_revise}{Comparison of convergence curves of the proposed GDPG algorithm and some state-of-the-art DRL algorithms. (a) Reward per episode (b) Mean sum channel capacity per step. (c) Mean fairness index per step. (d) Number of covert constraint violations per episode.}}}
    \label{Fig: Convergence Curves}
\end{figure*}
\par {\color{color_revise}{In order to make a fair performance comparison with benchmarks, the neural network architectures of all DRL-based benchmarks comprise two hidden layers for both the actor and critic networks, with each layer containing 256 neurons. In the proposed GDPG algorithm, the learning rates for its GDM and critic networks are used as the common settings for all other DRL-based benchmarks.}}


\subsection{Simulation Results}
\label{sub_sec_simulation_results}

\subsubsection{Comparison with Other Approaches}
\label{sub_sub_sec_comparison_with_Other Approaches}

\par We compare our approach with some other approaches in terms of the communication performance and covert of the system. As shown in Fig. \ref{Fig: Different Approaches}, NA approach stands out in terms of covert constraint compliance while remaining relatively low in terms of its mean sum channel capacity compared to other approaches with active transmissive RIS since the signal amplification is limited. On the other hand, our approach not only achieves higher communication performance but also maintains covert constraint compliance effectively. Unlike the approaches that only optimize certain aspects, such as OTO or OABO approaches, our approach integrates a holistic optimization strategy, thereby addressing both the communication performance and system covertness simultaneously. 

\subsubsection{Comparison with State-of-the-Art DRL Algorithms}
\label{sub_sub_sec_comparison_with_State-of-the-Art DRL Algorithms}

\par Fig. \ref{Fig: Convergence Curves}(a) illustrates the reward per episode of the proposed GDPG algorithm in comparison to other state-of-the-art DRL algorithms during the training process. As can be seen, the proposed GDPG algorithm consistently outperforms the other algorithms with a steady increase in reward over the training time. The success of the GDPG algorithm lies in the use of the GDM, which improves the exploration and policy optimization by generating the diverse and high-quality samples through its denoising process, thereby resulting in higher overall rewards, especially in the complex satellite-terrestrial covert communication scenarios. Moreover, DDPG and TD3 demonstrate initial improvements in rewards during the early stages of training, but their progress slows and eventually plateaus. This indicates that although they achieve early gains, their performance is constrained by limited exploration, thereby leading to suboptimal policy. Finally, SAC performs better than DDPG and TD3 with a more consistent increase in rewards. However, it still has lower performance than that of the GDPG algorithm, thereby illustrating that SAC lacks the same level of exploration efficiency and optimization capabilities provided by the proposed GDPG algorithm. {\color{color_revise}{In addition, the VAE-enabled DPG algorithm outperforms conventional baselines but underperforms the proposed GDPG algorithm, which confirms that variational autoencoders lack the multimodal expressiveness and stability of GDM in high-dimensional constrained spaces while improving exploration.}}

\par Fig. \ref{Fig: Convergence Curves}(b) shows the mean sum channel capacity per step of the proposed GDPG algorithm in comparison to other state-of-the-art DRL algorithms. The performance achieved by the proposed GDPG algorithm is notably higher than that of other algorithms. This indicates that the GDM-enabled DRL algorithm is more effective in leveraging its representation and exploration capabilities in the large state and action spaces. Another important metric is the fairness index, which is presented in Fig. \ref{Fig: Convergence Curves}(c). {\color{color_revise}{As can be seen, the GDPG and VAE-enabled DPG algorithms maintain the highest fairness indexes across all algorithms. However, the GDPG algorithm achieves the fastest convergence and consistently maintains a higher mean sum channel capacity per step, owing to its denoising process and more effective exploration of the action space.}}

\par Fig. \ref{Fig: Convergence Curves}(d) reveals the number of covert constraint violations per episode of the proposed GDPG algorithm in comparison to other algorithms. It is evident that the proposed GDPG algorithm effectively minimizes covert constraint violations while maintaining the lowest number of violations throughout the training phase. On the other hand, SAC results in a higher number of violations compared to the GDPG algorithm, especially in the early stages of training, which can be attributed to the entropy-driven exploration strategy that sometimes fails to maintain the strict covert constraints. Moreover, both DDPG and TD3 show similar trends, with DDPG experiencing a slightly higher number of violations than TD3. These algorithms struggle more than the GDPG algorithm in maintaining covert communication, likely due to their more limited exploration capabilities and lack of efficient optimization in complex environments. 

\begin{figure}[t]
	\centering
	\includegraphics[width=\linewidth]{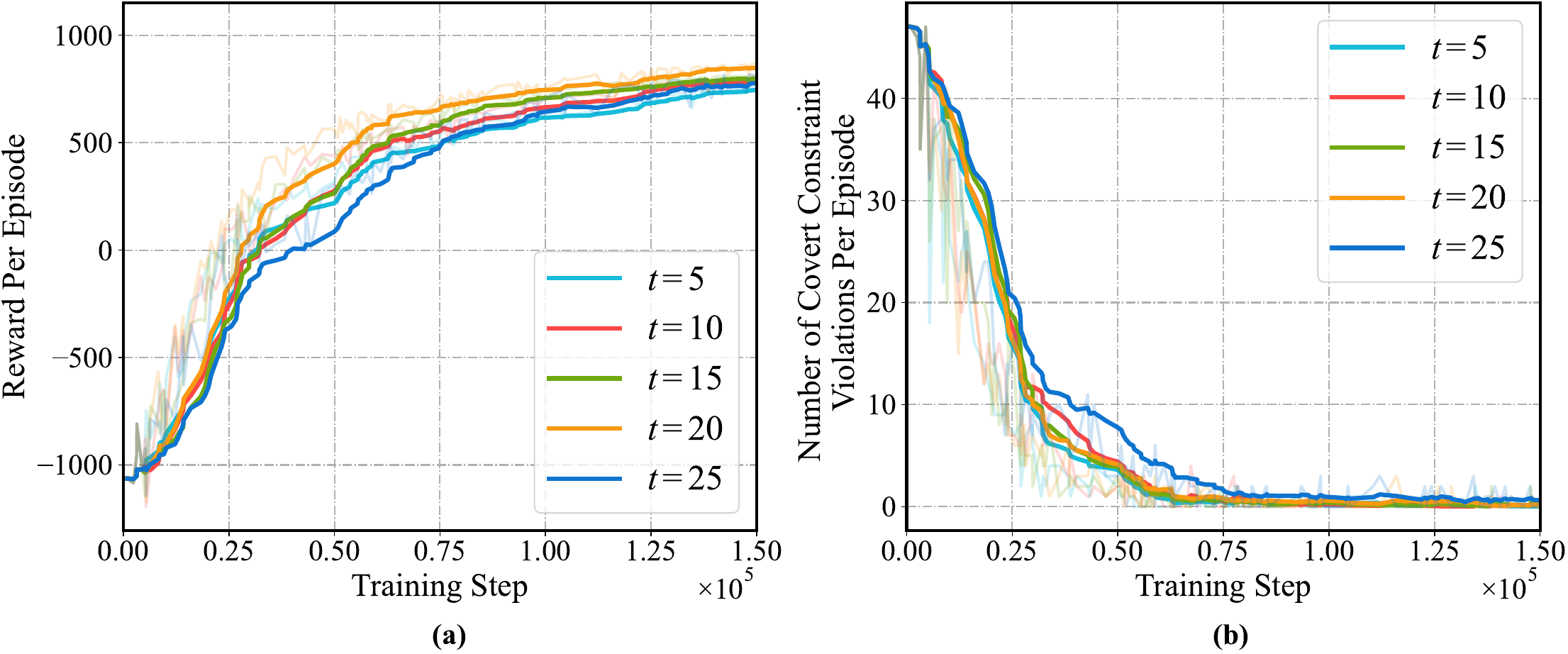}
	\caption{Performance evaluations on different diffusion steps $t$. (a) Reward per episode. (b) Number of covert constraint violations per episode.}
	\label{Fig: Diffusion Steps}
\vspace{-5mm}
\end{figure}

\subsubsection{Impact of Different Diffusion Steps}
\label{sub_sub_sec_impact_of_different_diffusion_steps}

\par Figs. \ref{Fig: Diffusion Steps}(a) and (b) show the performance of the proposed GDPG algorithm evaluated over different diffusion steps, specifically comparing the reward per episode and number of covert constraint violations per episode. To make the differences clearer, only the 150,000 training steps of the early training phase are shown. As can be seen in Fig. \ref{Fig: Diffusion Steps}(a), regardless of the number of denoising steps, the algorithm demonstrates relatively stable progress in terms of rewards. This indicates that different denoising steps can effectively boost the reward accumulation, and the performance remains stable to a certain extent throughout the training process.

\par However, despite the stable performance in terms of rewards, the number of denoising steps is not always better when increased or decreased. From the results of Fig. \ref{Fig: Diffusion Steps}(b), too many or few denoising steps can lead to slower convergence in terms of the number of covert constraint violations per episode. This suggests that the optimal number of denoising steps should strike a balance between the stability and convergence speed. Moreover, the results in the figure indicate that $t=10$ might be the optimal choice, as it achieves both relatively stable reward improvement and avoids excessively slow convergence.

\subsubsection{Impact of Different Learning Rates}
\label{sub_sub_sec_impact_of_different_learning_rates}

\par
\begin{figure}
    \centering
    \includegraphics[width=\linewidth]{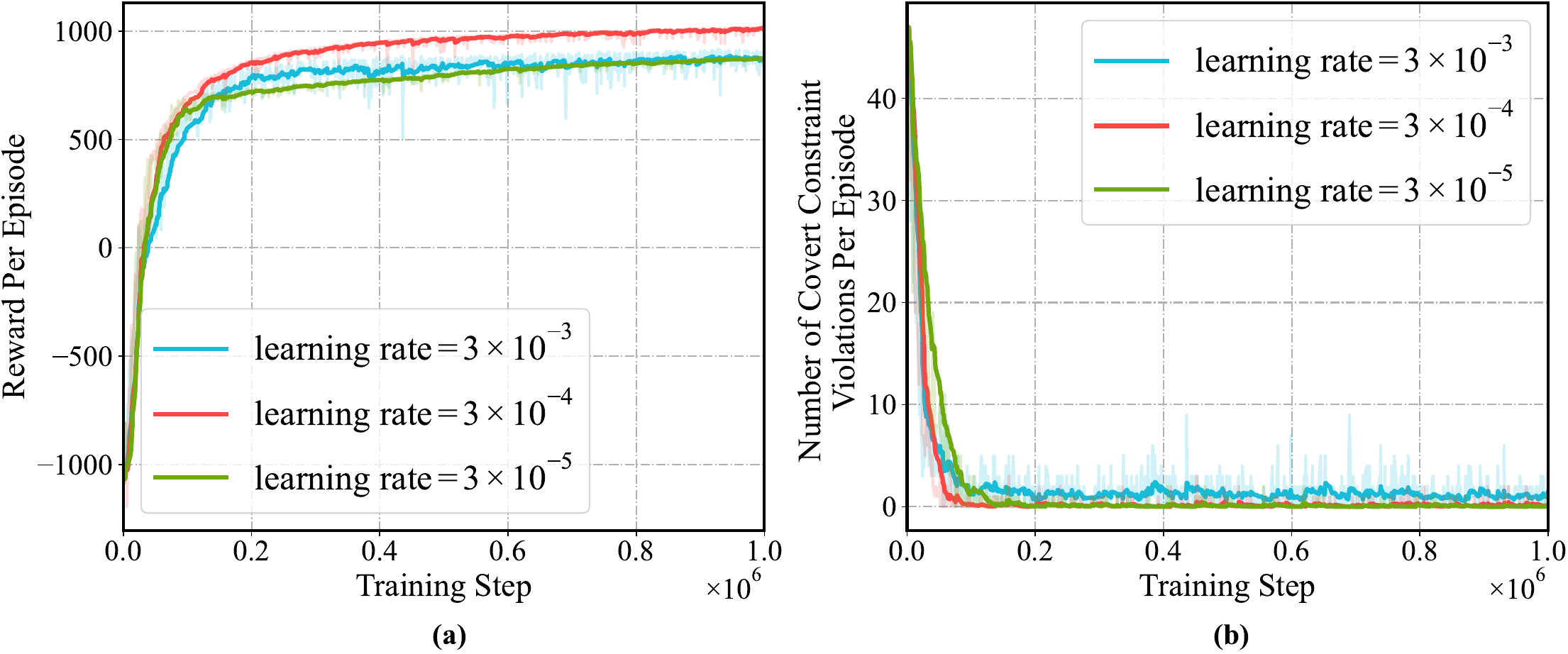}
    \caption{Performance evaluations on different learning rates. (a) Reward per episode. (b) Number of covert constraint violations per episode.}
    \label{Fig: Different_learning_rates}
\vspace{-5mm}
\end{figure}

 Fig. \ref{Fig: Different_learning_rates}(a) illustrates the impact of different learning rates on the convergence of reward per episode. It is observed that the reward increases rapidly at the initial training steps for all learning rates. Among them, the learning rate $3 \times 10^{-4}$ provides the best balance between the convergence speed and stability, achieving the highest and most stable final reward. Conversely, the learning rate $3 \times 10^{-3}$ exhibits faster initial learning but results in higher fluctuations, whereas a smaller learning rate leads to slow convergence with relatively lower performance. 

\par Fig. \ref{Fig: Different_learning_rates}(b) shows the number of covert constraint violations per episode for different learning rates during the training phase. Initially, all three cases exhibit a high number of violations, rapidly decreasing as training progresses. The learning rate $3 \times 10^{-4}$ achieves the fastest and most stable convergence towards nearly zero violations. In contrast, the largest learning rate shows significant oscillations, and the smallest learning rate converges slowly. Thus, a moderate learning rate is preferred for effectively balancing the convergence speed and system constraint compliance.

\subsubsection{Impact of Different Covert Requirements}
\label{sub_sub_sec_impact_of_different_covert_requirements}

\begin{figure}[t]
	\centering
	\includegraphics[width=\linewidth]{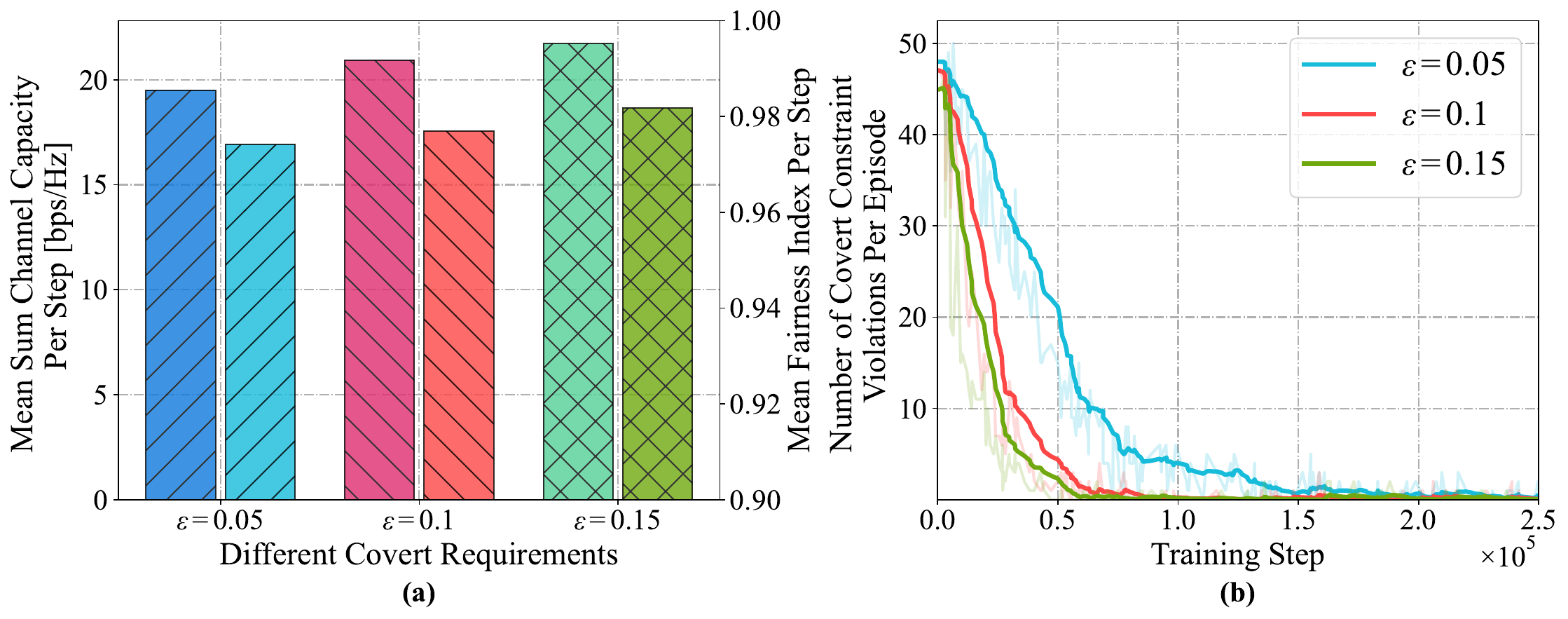}
	\caption{Performance evaluations on different covert requirements $\varepsilon$. (a) Mean sum channel capacity and mean fairness index per step. (b) Number of covert constraint violations per episode.}
	\label{Fig: Covert Requirement}
\vspace{-5mm}
\end{figure}

\par In Fig. \ref{Fig: Covert Requirement}(a), the mean sum channel capacity and fairness index per step are shown for different covert requirements. As the covert requirement becomes more stringent (i.e., $\varepsilon$ decreases), the mean sum channel capacity and mean fairness index per step slightly decrease. This indicates that enforcing higher covertness reduces the available channel capacity and fairness, thereby highlighting the trade-off between ensuring the security and maximizing the communication performance.

\par Fig. \ref{Fig: Covert Requirement}(b) shows the number of covert constraint violations per episode for various covert requirements. As expected, a stricter covert requirement results in a higher number of covert constraint violations, particularly at the early stage of training. This is due to the difficulty of adhering to more stringent covert constraints, which leads to more violations during the exploration phase. However, as the training progresses, the algorithm increasingly adheres to the covert constraint with the number of violations gradually approaching zero. 

\par In conclusion, the results show that while the GDPG algorithm can effectively maintain covertness with a relatively low number of violations, increasing the covert requirement introduces a trade-off between the sum channel capacity of all ground users, fairness among all ground users, and covertness of the considered system.

\subsubsection{Impact of Different AAT-RIS Sizes}
\label{sub_sub_sec_impact_of_different_AAT-RIS_sizes}

\begin{figure}[t]
	\centering
	\includegraphics[width=\linewidth]{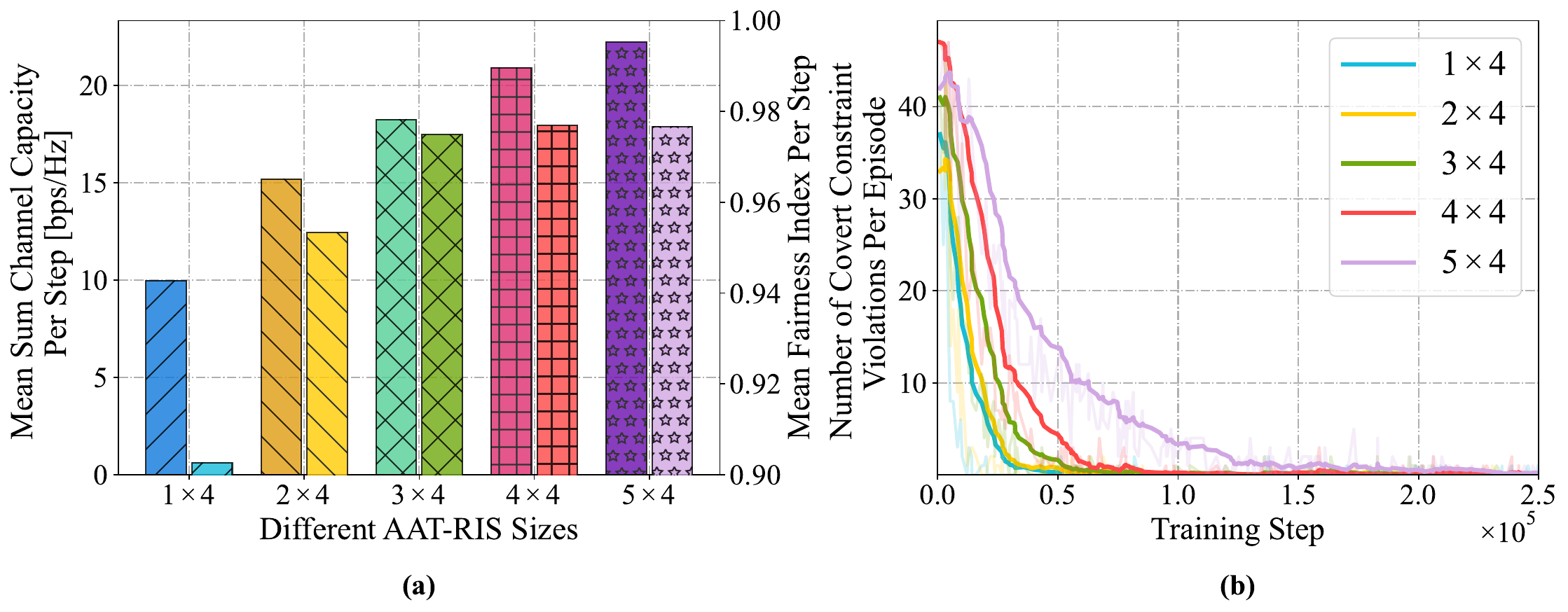}
	\caption{Performance evaluations on different AAT-RIS sizes $M_{x} \times M_{y}$. (a) Mean sum channel capacity and mean fairness index per step. (b) Number of covert constraint violations per episode.}
	\label{Fig: RIS Size}
\vspace{-5mm}
\end{figure}

\par Fig. \ref{Fig: RIS Size}(a) shows the performance of the proposed GDPG algorithm with different AAT-RIS sizes. As the AAT-RIS size increases, the mean sum channel capacity per step gradually improves. Specifically, larger configurations, such as $4 \times 4$ and $5 \times 4$, outperform smaller ones like $1 \times 4$ and $2 \times 4$. This improvement occurs because larger AAT-RIS sizes have more transmission and reflection elements, which results in the enhanced signal amplification and reconfiguration capabilities, thereby leading to better communication performance. Moreover, the mean fairness index per step shows a similar trend. Specifically, larger AAT-RIS sizes provide better fairness, which suggests that the larger configurations help balance the communication capacity among multiple users more effectively. However, the performance gains become less pronounced as the size increases, which indicates that the optimal configuration may depend on a balance between the sum channel capacity of all users, fairness among all users and computational complexity.

\par Fig. \ref{Fig: RIS Size}(b) presents the number of covert constraint violations per episode during the early training phase for different AAT-RIS sizes. Specifically, smaller AAT-RIS sizes, such as $1 \times 4$, result in fewer covert constraint violations earlier in the training process compared to larger sizes like $4 \times 4$ and $5 \times 4$. This happens because smaller configurations introduce less complexity, allowing the algorithm to meet the covert communication requirements more quickly. While the larger AAT-RIS sizes provide higher channel capacity, they also introduce more complexity in satisfying the covert constraints, which results in slower convergence during the early stages of training. Therefore, smaller AAT-RIS sizes enable faster convergence with fewer covert constraint violations, while larger sizes require more time to optimize communication parameters and satisfy the covert requirements.


\section{Conclusion}
\label{sec_conclusion}

\par This paper has investigated a novel AAT-RIS assisted satellite-terrestrial covert communication scheme against a ground Warden. To achieve better communication performance and greater fairness among ground users while ensuring covertness, we have derived the covert constraint and formulated the ASCCOP to maximize the sum channel capacity of all ground users by jointly optimizing the trajectory and active beamforming of the AAT-RIS. To deal with the long-term optimization problem, we have reformulated it into the framework of DRL and proposed a GDPG algorithm that leverages the GDM as policy representation to produce multimodal actions and adopts the action gradient method to improve the GDM-based policy. The simulation results have demonstrated the effectiveness of our approach over several baseline approaches. Moreover, we have also verified the robustness of the GDPG algorithm under different environment settings and algorithm parameters. In future work, we will further study the joint implementation of LEO satellites and AAT-RIS to improve covertness performance.


\ifCLASSOPTIONcaptionsoff
\newpage
\fi

\bibliographystyle{IEEEtran}
\bibliography{references.bib}

@Article{Mahboob2024,
  author    = {Shadab Mahboob and Lingjia Liu},
  journal   = {{IEEE} Commun. Surv. Tutorials},
  title     = {Revolutionizing Future Connectivity: {A} Contemporary Survey on {AI}-Empowered Satellite-Based Non-Terrestrial Networks in {6G}},
  year      = {2024},
  month ={2nd Quart.,},
  number    = {2},
  pages     = {1279--1321},
  volume    = {26},
}

@Article{Zheng2022,
  author    = {Beixiong Zheng and Shaoe Lin and Rui Zhang},
  journal   = {{IEEE} J. Sel. Areas Commun.},
  title     = {Intelligent Reflecting Surface-Aided {LEO} Satellite Communication: Cooperative Passive Beamforming and Distributed Channel Estimation},
  year      = {2022},
  month     = {Oct.},
  number    = {10},
  pages     = {3057--3070},
  volume    = {40},
}

@ARTICLE{Lizhen2024,
  author={Li, Zhendong and Chen, Wen and Wu, Qingqing and Zhu, Xusheng and Qin, Haoran and Wang, Kunlun and Li, Jun},
  journal={{IEEE} Internet Things J.}, 
  title={Toward Transmissive {RIS} Transceiver Enabled Uplink Communication Systems: Design and Optimization}, 
  year={2024},
  volume={11},
  number={4},
  month={Feb.},
  pages={6788-6801},
}

@Article{Guo2025,
  author    = {Jiajia Guo and Xi Yang and Chao{-}Kai Wen and Shi Jin and Geoffrey Ye Li},
  journal   = {{IEEE} Trans. Commun.},
  title     = {Deep Learning-Based {CSI} Feedback for {RIS}-Assisted Multi-User Systems},
  year      = {2025},
  month     = {Jul.},
  number    = {7},
  pages     = {4974--4989},
  volume    = {73},
}

@ARTICLE{Kang2025,
  author={Kang, Bichen and Ye, Neng and An, Jianping},
  journal={{IEEE} Trans. Inf. Forensics Secur.}, 
  title={Achieving Positive Rate of Covert Communications Covered by Randomly Activated Overt Users}, 
  year={2025},
  month={Feb.},
  volume={20},
  number={},
  pages={2480-2495},
}

@ARTICLE{WangMan2025,
  author={Wang, Manlin and Yao, Yao and Ding, Haiyang and Shao, Shihai and Xia, Bin and Wang, Jiangzhou},
  journal={{IEEE} Trans. Inf. Forensics Secur.}, 
  title={Angle and Distance Discrimination by Utilizing Frequency Conversion Capability of STC-IRS for Covert Communications}, 
  year={2025},
  month= {Jan.},
  volume={20},
  number={},
  pages={1503-1518},
}

@Article{Shahzad2018,
  author    = {Khurram Shahzad and Xiangyun Zhou and Shihao Yan and Jinsong Hu and Feng Shu and Jun Li},
  journal   = {{IEEE} Trans. Wirel. Commun.},
  title     = {Achieving Covert Wireless Communications Using a Full-Duplex Receiver},
  year      = {2018},
  month     = {Dec.},
  number    = {12},
  pages     = {8517--8530},
  volume    = {17},
}

@Article{He2025,
  author    = {Rongrong He and Guoxin Li and Jin Chen and Haichao Wang and Xinrong Guan and Yifan Xu and Wenhui He and Yuhua Xu},
  journal   = {{IEEE} Trans. Wirel. Commun.},
  title     = {Joint Power and Beamformer Optimization in Multi-Antenna Relay Covert System: Exploiting Public Users as Shelter},
  year      = {2025},
  month     = {Jan.},
  number    = {1},
  pages     = {385--400},
  volume    = {24},
}

@InProceedings{R2-3-1,
  author    = {Romain Lopez and Pierre Boyeau and Nir Yosef and Michael I. Jordan and Jeffrey Regier},
  booktitle = {Proc. Adv. Neural Inf. Process Syst. 33, NeurIPS 2020},
  title     = {Decision-Making with Auto-Encoding Variational Bayes},
  year      = {2020},
  month     = {Dec. 6-12,},
  location = {Virtual},
}

@Article{Zhang2025,
  author  = {Chuang Zhang and Geng Sun and Jiahui Li and Qingqing Wu and Jiacheng Wang and Dusit Niyato and Yuanwei Liu},
  journal = {{IEEE} Trans. Mob. Comput.},
  title   = {Multi-Objective Aerial Collaborative Secure Communication Optimization via Generative Diffusion Model-Enabled Deep Reinforcement Learning},
  year    = {2025},
  month   = {Apr.},
  number  = {4},
  pages   = {3041--3058},
  volume  = {24},
}

@Article{Li2024,
  author  = {Jiahui Li and Geng Sun and Qingqing Wu and Dusit Niyato and Jiawen Kang and Abbas Jamalipour and Victor C. M. Leung},
  journal = {{IEEE} J. Sel. Areas Commun.},
  title   = {Collaborative Ground-Space Communications via Evolutionary Multi-Objective Deep Reinforcement Learning},
  year    = {2024},
  month   = {Dec.},
  number  = {12},
  pages   = {3395--3411},
  volume  = {42},
}

@Article{AlHourani2020,
  author  = {Akram Al{-}Hourani},
  journal = {{IEEE} Wirel. Commun. Lett.},
  title   = {On the Probability of {L}ine-of-{S}ight in Urban Environments},
  year    = {2020},
  month = {Aug.},
  number  = {8},
  pages   = {1178--1181},
  volume  = {9},
}

@InProceedings{Zhanglei2024,
  author    = {Zhang, Lei and Chen, Zhao and Jiang, Chunxiao and Yin, Liuguo},
  booktitle = {Proc. IEEE Int. Conf. Commun. (ICC)},
  title     = {Covert Communication in Ultra-Dense {LEO} Satellite Systems with Interference Uncertainty},
  year      = {2024},
  address   = {Denver, CO, USA},
  month     = {Jun. 9-13},
  pages     = {1255-1260},
}

@InProceedings{Hui2024,
  author    = {Pei Hui and Lei Guan and Zan Li and Chenxi Li and Wendong Gao and Hanwen Zhang},
  booktitle = {Proc. IEEE Int. Conf. Commun. (ICC)},
  title     = {Covert Transmission Control Scheme for Terrestrial-satellite Communications},
  year      = {2024},
  address   = {Denver, CO, USA},
  month     = {Jun. 9-13},
  pages     = {1974--1979},
}

@Article{Song2023,
  author    = {Da Song and Ziyi Yang and Gaofeng Pan and Shuai Wang and Jianping An},
  journal   = {{IEEE} Internet Things J.},
  title     = {{RIS}-Assisted Covert Transmission in Satellite-Terrestrial Communication Systems},
  year      = {2023},
  month = {Nov.},
  number    = {22},
  pages     = {19415--19426},
  volume    = {10},
}

@Article{Ding2024,
  author    = {Yu Ding and Qingqing Zhang and Weidang Lu and Nan Zhao and Arumugam Nallanathan and Xianbin Wang and Xiaoniu Yang},
  journal   = {{IEEE} Trans. Wirel. Commun.},
  title     = {Collaborative Communication and Computation for Secure {UAV}-Enabled {MEC} Against Active Aerial Eavesdropping},
  year      = {2024},
  month = {Nov.},
  number    = {11},
  pages     = {15915--15929},
  volume    = {23},
}

@Article{WDLukito2024,
  author  = {Lukito, William D. and Xiang, Wei and Lai, Phu and Cheng, Peng and Liu, Chang and Yu, Kan and Zhu, Xiaoyan},
  journal = {{IEEE} Internet Things J.},
  title   = {Integrated {STAR-RIS} and {UAV} for Satellite {IoT} Communications: An Energy-Efficient Approach},
  year    = {2025},
  month = {May},
  volume={12},
  number={9},
  pages={11356--11371},
}

@Article{Feng2024,
  author  = {Feng, Kai and Zhou, Ting and Xu, Tianheng and Chen, Xianfu and Hu, Honglin and Wu, Celimuge},
  journal = {{IEEE} Internet Things J.},
  title   = {Reconfigurable Intelligent Surface-Assisted Multisatellite Cooperative Downlink Beamforming},
  year={2024},
  month = {Jul.},
  volume={11},
  number={13},
  pages={23222-23235},
}

@Article{Wang2025,
  author  = {Wang, Qunshu and Guo, Shaoyong and Wu, Celimuge and Xing, Chengwen and Zhao, Nan and Niyato, Dusit and Karagiannidis, George K.},
  journal = {{IEEE} J. Sel. Areas Commun.},
  title   = {{STAR-RIS} Aided Covert Communication in {UAV} Air-Ground Networks},
  year      = {2025},
  month = {Jan.},
  number    = {1},
  pages     = {245-259},
  volume    = {43},
}

@Article{Lu2021,
  author  = {Haiquan Lu and Yong Zeng and Shi Jin and Rui Zhang},
  journal = {{IEEE} Trans. Wirel. Commun.},
  title   = {Aerial Intelligent Reflecting Surface: Joint Placement and Passive Beamforming Design With {3D} Beam Flattening},
  year    = {2021},
  month   = {Jul.},
  number  = {7},
  pages   = {4128--4143},
  volume  = {20},
}

@Article{Wang2023,
  author  = {Chao Wang and Xinying Chen and Jianping An and Zehui Xiong and Chengwen Xing and Nan Zhao and Dusit Niyato},
  journal = {{IEEE} Trans. Commun.},
  title   = {Covert Communication Assisted by {UAV-IRS}},
  year    = {2023},
  month   = {Jan.},
  number  = {1},
  pages   = {357--369},
  volume  = {71},
}

@inproceedings{li2024two,
  title={Two-Way Aerial Secure Communications via Distributed Collaborative Beamforming under Eavesdropper Collusion},
  author={Li, Jiahui and Sun, Geng and Wu, Qingqing and Liang, Shuang and Wang, Pengfei and Niyato, Dusit},
  booktitle={Proc. IEEE Conf. Comput. Commun. (INFOCOM)},
  pages={331--340},
  month = {May 20-23,},
  address = {Vancouver, Canada},
  year={2024},
}

@Article{Chen2023,
  author  = {Xinying Chen and Jianping An and Zehui Xiong and Chengwen Xing and Nan Zhao and F. Richard Yu and Arumugam Nallanathan},
  journal = {{IEEE} Commun. Surv. Tutorials},
  title   = {Covert Communications: {A} Comprehensive Survey},
  year    = {2023},
  month   = {2nd Quart.,},
  number  = {2},
  pages   = {1173--1198},
  volume  = {25},
}

@InProceedings{Wu2022,
  author  = {Zeke Wu and Haifeng Shuai and Rui Liu and Kefeng Guo and Shibing Zhu},
  title   = {Performance Analysis of Covert Communication Based on Integrated Satellite Multiple Terrestrial Relay Networks},
  year    = {2022},
  address = {Chengdu, China},
  pages   = {380--385},
  month    = {Dec. 9-12},
  booktitle = {Proc. 2022 {IEEE} 8th Int. Conf. Comput. Commun. (ICCC)},
}

@Article{Jia2025,
  author  = {Jia, Huaiqi and Wang, Ying and Wu, Wen and Yuan, Jun},
  journal = {{IEEE} Internet Things J.},
  title   = {Robust Transmission Design for Covert Satellite Communication Systems With Dual-{CSI} Uncertainty},
  year    = {2025},
  month = {Jun.},
  volume={12},
  number={12},
  pages={21892--21903},
}

@Article{Yu2024,
  author  = {Yu, Hanpeng and Yu, Jihong and Liu, Jiahao and Li, Yun and Ye, Neng and Yang, Kai and An, Jianping},
  journal = {{IEEE} Trans. Aerosp. Electron. Syst.},
  title   = {Covert Satellite Communication Over Overt Channel: A Randomized Gaussian Signalling Approach},
  year    = {2025},
  month   = {Apr.},
  number  = {2},
  pages   = {2355--2368},
  volume  = {61},

}

@Article{Zheng2025,
  author  = {Zheng, Ziyuan and Jing, Wenpeng and Lu, Zhaoming and Wu, Qingqing and Zhang, Haijun and Gesbert, David},
  journal = {{IEEE} J. Sel. Areas Commun.},
  title   = {Cooperative Multi-Satellite and Multi-{RIS} Beamforming: Enhancing {LEO} {SatCom} and Mitigating {LEO-GEO} Intersystem Interference},
  year    = {2025},
  month   = {Jan.},
  number  = {1},
  pages   = {279--296},
  volume  = {43},
}

@Article{Zhang2024,
  author  = {Jifa Zhang and Wei Wang and Yuan Gao and Weidang Lu and Nan Zhao and Dusit Niyato},
  journal = {{IEEE} Trans. Wirel. Commun.},
  title   = {Robust Covert Multicasting Aided by {STAR-RIS} With Hardware Impairment},
  year    = {2024},
  month   = {Nov.},
  number  = {11},
  pages   = {16172--16186},
  volume  = {23},
}

@InProceedings{Cao2024,
  author    = {Yizhi Cao and Zewei Guo and Qifeng Miao and Ranran Sun and Ji He and Xiaochen Li},
  booktitle = {Proc. Int. Conf. Networking Network Appl. (NaNA)},
  title     = {Minimization of Age of Information for Satellite-Terrestrial Covert Communication with a Full-Duplex Receiver},
  year      = {2024},
  month = {Aug. 9-12,},
  address = {Yinchuan City, China},
  pages     = {242--246},
}

@InProceedings{Guo2024,
  author          = {Zewei Guo and Ranran Sun and Ji He and Yulong Shen and Xiaohong Jiang},
  title           = {Covert Communication in Satellite-Terrestrial Systems with a Full-Duplex Receiver},
  year            = {2024},
  address         = {Xi'an City, China},
  pages           = {72--77},
  month            = {Oct. 25-27,},
  booktitle         = {Proc. 2024 Int. Conf. Satell. Internet (SAT-NET)},
}

@Article{Sediq2013,
  author  = {Akram Bin Sediq and Ramy H. Gohary and Rainer Schoenen and Halim Yanikomeroglu},
  journal = {{IEEE} Trans. Wirel. Commun.},
  title   = {Optimal Tradeoff Between Sum-Rate Efficiency and {Jain}'s Fairness Index in Resource Allocation},
  year    = {2013},
  month   = {Jul.},
  number  = {7},
  pages   = {3496--3509},
  volume  = {12},
}

@Article{Yimengjie2025,
    author = {Mengjie Yi and Xijun Wang and Juan Liu and Yan Zhang and Ronghui Hou},
    journal = {{IEEE} Trans. Commun.},
    title = {Meta-Reinforcement Learning for Timely and Energy-efficient Data Collection in Solar-powered {UAV}-assisted {IoT} Networks},
    year    = {Early Access, 2025},
    note    = {doi: {10.1109/TCOMM.2025.3543185}},
}

@InProceedings{Wang2023a,
  author    = {Zhendong Wang and Jonathan J. Hunt and Mingyuan Zhou},
  booktitle = {Proc. 11th Int. Conf. Learn. Representations (ICLR)},
  title     = {Diffusion Policies as an Expressive Policy Class for Offline Reinforcement Learning},
  year      = {2023},
  address   = {Kigali, Rwanda},
  month     = {May 1-5,},
  pages     = {1--17},
}

@ARTICLE{SongR2024,
  author={Song, Rongguang and Yin, Haifan and Wang, Zipeng and Yang, Taorui and Ren, Xue},
  journal={{IEEE} Trans. Antennas Propag.}, 
  title={Modeling, Design, and Verification of an Active Transmissive {RIS}}, 
  year={2024},
  month = {Dec.},
  volume={72},
  number={12},
  pages={9239-9250},
}

@InProceedings{Silver2014,
  author    = {David Silver and Guy Lever and Nicolas Heess and Thomas Degris and Daan Wierstra and Martin A. Riedmiller},
  booktitle = {Proc. 31st Int. Conf. Mach. Learn. (ICML)},
  title     = {Deterministic Policy Gradient Algorithms},
  year      = {2014},
  month     = {Jun. 21-25},
  address   = {Beijing, China},
  pages     = {387--395},
}

@Article{Jiahui2025,
  author        = {Li, Jiahui and Sun, Geng and Wu, Qingqing and Liang, Shuang and Wang, Jiacheng and Niyato, Dusit and Kim, Dong In},
  title         = {Aerial Secure Collaborative Communications under Eavesdropper Collusion in Low-altitude Economy: A Generative Swarm Intelligent Approach},
  year          = {2025},
  month         = {Mar.},
  journal = {arXiv preprint arXiv:2503.00721},
  note = {doi: {10.48550/arXiv.2503.00721}},
}

@Article{Yang2023,
  author        = {Long Yang and Zhixiong Huang and Fenghao Lei and Yucun Zhong and Yiming Yang and Cong Fang and Shiting Wen and Binbin Zhou and Zhouchen Lin},
  title         = {Policy Representation via Diffusion Probability Model for Reinforcement Learning},
  year          = {2023},
  month         = {May},
  journal = {arXiv preprint arXiv:2305.13122},
  note = {doi: {10.48550/arXiv.2305.13122}},
}

@InProceedings{Dhariwal2021,
  author    = {Prafulla Dhariwal and Alexander Quinn Nichol},
  booktitle = {Proc. Adv. Neural Inf. Process. Syst. 2021 (NIPS)},
  title     = {Diffusion Models Beat {GANs} on Image Synthesis},
  year      = {2021},
  address   = {virtual},
  month     = {Dec. 6-14},
  pages     = {8780--8794},
}

@Article{Haarnoja2018,
  author        = {Tuomas Haarnoja and Aurick Zhou and Kristian Hartikainen and George Tucker and Sehoon Ha and Jie Tan and Vikash Kumar and Henry Zhu and Abhishek Gupta and Pieter Abbeel and Sergey Levine},
  journal       = {arXiv preprint 		arXiv:1812.05905},
  title         = {Soft Actor-Critic Algorithms and Applications},
  year          = {2018},
  month = {Dec.},
  note = {doi:{10.48550/arXiv.1812.05905}},
}

@InProceedings{Fujimoto2018,
  author    = {Scott Fujimoto and Herke van Hoof and David Meger},
  booktitle = {Proc. 35th Int. Conf. Mach. Learn. (ICML)},
  title     = {Addressing Function Approximation Error in Actor-Critic Methods},
  year      = {2018},
  month = {Jul. 10-15,},
  address = {Stockholmsm{\"{a}}ssan, Stockholm, Sweden},
  pages     = {1582--1591},
}

@Article{Hieu2023,
  author    = {Nguyen Quang Hieu and Dinh Thai Hoang and Dusit Niyato and Diep N. Nguyen and Dong In Kim and Abbas Jamalipour},
  journal   = {{IEEE} Trans. Commun.},
  title     = {Joint Power Allocation and Rate Control for Rate Splitting Multiple Access Networks With Covert Communications},
  year      = {2023},
month = {Apr.},
  number    = {4},
  pages     = {2274--2287},
  volume    = {71},
}

@Article{Hu2021,
  author    = {Shuyan Hu and Wei Ni and Xin Wang and Abbas Jamalipour and Dean Ta},
  journal   = {{IEEE} Trans. Inf. Forensics Secur.},
  title     = {Joint Optimization of Trajectory, Propulsion, and Thrust Powers for Covert {UAV-on-UAV} Video Tracking and Surveillance},
  year      = {2021},
  pages     = {1959--1972},
  volume    = {16},

}

@Article{Xing2024,
  author  = {Jintao Xing and Tiejun Lv and Weicai Li and Wei Ni and Abbas Jamalipour},
  journal = {{IEEE} Internet Things J.},
  title   = {Joint Optimization of Beamforming and Noise Injection for Covert Downlink Transmissions in Cell-Free Internet of Things Networks},
  year    = {2024},
  month   = {Mar.},
  number  = {6},
  pages   = {10525--10536},
  volume  = {11},
}

\vfill

\end{document}